\documentclass[a4paper,fleqn,usenatbib]{mnras}

\usepackage[T1]{fontenc}
\usepackage{ae,aecompl}

\usepackage{subfigure}
\usepackage{graphicx}	
\usepackage{amsmath}	
\usepackage{amssymb}	
\usepackage{color}
\usepackage{mathptmx}
\usepackage{epsfig}
\usepackage{wasysym}
\usepackage{xfrac}
\usepackage{pifont}
\usepackage[table,xcdraw]{xcolor}
\usepackage{booktabs}
\usepackage{multirow}
\usepackage{comment}
\usepackage{soul}
\usepackage{tikz}
\usepackage{booktabs}
\usepackage{orcidlink}

\definecolor{grey}{rgb}{0.75,0.75,0.75}
\definecolor{Orange}{rgb}{1.0,0.5,0.15}
\definecolor{brown}{rgb}{0.7,0.25,0.0}
\definecolor{pink}{rgb}{1.0,0.5,0.5}
\definecolor{darkerred}{rgb}{0,0.5,0.5}
\definecolor{darkerblue}{rgb}{0,0,0.8}
\definecolor{lightblue}{rgb}{0.12, 0.56, 1.0}
\definecolor{Blue}{rgb}{0,0.08,0.65}
\definecolor{Red}{rgb}{0.65,0.08,0.05}
\definecolor{Green}{rgb}{0.15,0.45,0.25}

\definecolor{purple}{rgb}{0.5,0,0.87}

\usetikzlibrary{positioning,decorations.pathreplacing,shapes}

\title[Structural diversity in low-mass galaxies]{Cosmic reflections I: the structural diversity of simulated and observed low-mass galaxy analogues}

\author[G. Martin et al.]{G. Martin\orcidlink{0000-0003-2939-8668},$^{1,2,3}$\thanks{E-mail: garreth.martin@nottingham.ac.uk}
A. E. Watkins\orcidlink{0000-0003-4859-3290}$^{4}$,
Y. Dubois\orcidlink{0000-0003-0225-6387}$^{5}$,
J. Devriendt\orcidlink{0000-0002-8140-0422}$^{6}$,
S. Kaviraj\orcidlink{0000-0002-5601-575X}$^{4}$,
D. Kim\orcidlink{0000-0001-5120-0158}$^{7}$,
K. Kraljic\orcidlink{0000-0001-6180-0245}$^{8}$,\newauthor
I. Lazar\orcidlink{0009-0000-1797-0300}$^{4}$,
F. R. Pearce\orcidlink{0000-0002-2383-9250}$^{1}$,
S. Peirani\orcidlink{0000-0001-6902-2898}$^{9,10}$,
C. Pichon\orcidlink{0000-0003-0695-6735}$^{5,11}$,
A. Slyz$^{6}$,
S. K. Yi\orcidlink{0000-0002-4556-2619}$^{12}$
\\
$^{1}$School of Physics and Astronomy, University of Nottingham, University Park, Nottingham NG7 2RD, UK\\
$^{2}$Korea Astronomy and Space Science Institute, 776 Daedeokdae-ro, Yuseong-gu, Daejeon 34055, Korea\\
$^{3}$Steward Observatory, University of Arizona, 933 N. Cherry Ave, Tucson, AZ 85719, USA\\
$^{4}$Centre for Astrophysics Research, University of Hertfordshire, College Lane, Hatfield AL10 9AB, UK \\
$^{5}$Institut d’Astrophysique de Paris, UMR 7095, CNRS, Sorbonne Universit\'e, 98 bis boulevard Arago, 75014 Paris, France\\
$^{6}$Department of Physics, University of Oxford, Denys Wilkinson Building, Keble Road, Oxford OX1 3RH, UK\\
$^{7}$Astronomy and Space Science Department, Chungnam National University, Daehak-ro 99, Yuseong-gu Daejeon 34134, Republic of Korea\\
$^{8}$ Observatoire Astronomique de Strasbourg, Universit\'e de Strasbourg, CNRS, UMR 7550, F-67000 Strasbourg, France\\
$^{9}$ ILANCE, CNRS, University of Tokyo International Research Laboratory, Kashiwa, Chiba 277-8582, Japan\\
$^{10}$ Kavli IPMU (WPI), UTIAS, The University of Tokyo, Kashiwa, Chiba 277-8583, Japan\\
$^{11}$ Kyung Hee University, Dept. of Astronomy \& Space Science, Yongin-shi, Gyeonggi-do 17104, Republic of Korea\\
$^{12}$  Department of Astronomy, Yonsei University, 50 Yonsei-ro, Seodaemun-gu, Seoul 03722, Republic of Korea
}

\pubyear{2024}

\begin{document}
\label{firstpage}
\pagerange{\pageref{firstpage}--\pageref{lastpage}}
\maketitle


\begin{abstract}

\noindent Dwarf galaxies serve as powerful laboratories for investigating the underlying physics of galaxy evolution including the impact of baryonic feedback processes and external environmental influences. We compare the visual and structural properties of dwarf galaxies in ultra-deep HSC-SSP imaging of the COSMOS field with those measured from realistic HSC-like synthetic observations of dwarfs generated by the \textsc{Illustris TNG50} and \textsc{NewHorizon} simulations. Using S\'ersic  profile fitting and non-parametric morphological metrics (Gini, $M_{20}$, asymmetry, and concentration), we evaluate the diversity of structural properties in observed and simulated galaxies.
Our analysis shows that \textsc{NewHorizon} and \textsc{TNG50} galaxies lie at opposite extremes of observed structural trends: \textsc{NewHorizon} produces diffuse, extended galaxies with shallow S\'ersic  indices, while TNG50 yields compact, concentrated systems with steep indices. Both simulations reproduce observed structural trends more closely at higher stellar masses ($M_{\star} \sim 10^{9.5}~{\rm M_{\odot}}$) but fail to capture the full diversity of COSMOS dwarfs at lower masses. Non-parametric metrics further show that \textsc{NewHorizon} galaxies exhibit more uneven, clumpy light distributions while TNG50 galaxies have smoother but excessively concentrated profiles. These structural differences reflect underlying differences in their physical prescriptions and are likely driven by differing approaches to ISM gas physics, supernova feedback and star formation in addition to differences in numerical resolution.
Our findings highlight the unique power of low-mass galaxies to constrain differences in simulation physics, especially star formation and feedback prescriptions. Upcoming surveys from facilities like the Vera C. Rubin Observatory and Euclid will enable more rigorous comparisons with simulations, offering deeper insights into the physical processes shaping galaxy evolution.

\end{abstract}
\begin{keywords}
Galaxies: dwarf, structure, Methods: observational, numerical, data analysis
\end{keywords}



\section{Introduction}
\label{sec:introduction}

The study of galaxy evolution has made significant advances in recent decades, largely propelled by extensive wide-area surveys such as the Sloan Digital Sky Survey \citep[SDSS;][]{York2000}. A notable limitation of these surveys is their primary focus on detecting the brightest and most massive objects in the sky, with completeness rapidly decreasing for objects with surface brightnesses fainter than $\mu_{r} < 23$~mag~arcsec$^{-2}$ \citep[][]{Kniazev2004,Blanton2005,Driver2005}. Both observational and theoretical investigations \citep[e.g.][]{Disney1976,Dalcanton1997,Martin2019,Jackson2021,Kim2022} have postulated that they may only capture a fraction of the entire galaxy population, leaving us with an incomplete picture of galaxies and their evolution. 

Notably absent from past wide-area surveys are dwarf galaxies (especially at distances beyond the local Universe) as well as more massive objects like the giant low-surface-brightness (LSB) galaxies Malin 1 \citep{Bothun1987} and UGC 1382 \citep{Hagen2016}, as well as faint tidal structures found in the outskirts of galaxies and clusters \citep{Kaviraj2014b,Montes2021,Martin2022}.

This issue is likely to be particularly pronounced in the low-mass regime ($M_{\star}<10^{9.5}~{\rm M_{\odot}}$), where galaxies are intrinsically faint and are often associated with relatively intense and intermittent bursts of star formation \citep{Searle1973,Guo2016}, which introduces a degree of stochasticity to the identification of low-mass galaxies with otherwise similar properties in shallower surveys \citep[e.g.][]{Jackson2021,Kaviraj2025}. Furthermore, the shallower potential wells of these galaxies make them more vulnerable to processes that can result in them becoming increasingly diffuse \citep[e.g.][]{DiCintio2017,Martin2019}.

The detection of significant numbers of galaxies at the extremely faint and diffuse end of the low-surface-brightness population, now commonly referred to as 'ultra-diffuse galaxies' \citep[][UDGs]{vanDokkum2015}, underscores the presence of a notable bias in our current ability to accurately sample and understand low-mass galaxy populations \citep[e.g.][]{MartinezDelgado2016,Roman2019,Prole2021}. It is clear therefore that shallow imaging can introduce substantial biases in terms of which dwarf galaxy populations are recovered, limiting our understanding of galaxy populations and their evolution to a biased subset of the Universe.



Recent advancements have led to remarkable improvements in the sensitivity of wide-area survey instruments, enabling the routine detection of distant dwarf galaxies in various cosmic environments. Surveys now being conducted by next-generation ground- and space-based instruments including those undertaken by the Vera C. Rubin Observatory \citep[][]{Olivier2008,Ivezic2019},  JWST \citep{Gardner2006} and Euclid \citep{Laureijs2011,Borlaff2021} are poised to revolutionise galaxy evolution studies. These surveys will provide an unprecedented opportunity to gain a detailed statistical understanding of the LSB Universe. In a matter of days, the Rubin Observatory will surpass the depth and fidelity of the SDSS's initial 8-year survey in the Southern Sky. Moreover, it will continue to produce even deeper imaging over the next decade, achieving a maximum limiting surface brightness of $\mu_{r}\sim 30.3$~mag~arcsec$^{-2}$. Simultaneously, space-based, diffraction-limited observatories are already providing high-resolution observations of low-mass galaxies, reaching back as far as the early Universe \citep[e.g. COSMOS-Web, PANORAMIC;][]{Casey2022,Williams2021}. These new observational datasets will be transformative, enabling the study of large, unbiased galaxy samples across the low- and high-mass regimes and significantly advancing our statistical understanding of galaxy populations.

In parallel, high-resolution cosmological simulations are beginning to play a crucial role in guiding our exploration of poorly understood corners of discovery space. By facilitating realistic synthetic observations through forward modelling of simulation data within realistic cosmological contexts, these simulations offer comprehensive predictions with fidelity matching the capabilities of new instruments. Although simulations with sufficient resolution to investigate the resolved properties of dwarf galaxies have traditionally been confined to single-halo zoom-ins \citep[e.g.][]{Guedes2011,Hopkins2014,Wetzel2016}, larger simulations with box sizes spanning tens of Mpc are becoming more common \citep[e.g.][]{Tremmel2017,Nelson2019,Pillepich2019,Dubois2021,Feldmann2023}.

By simulating relatively large contiguous volumes, it is possible to overcome several shortcomings associated with zoom simulations, including limited sample sizes, potential selection biases, and a restricted ability to probe cosmic structures and correlations over large scales. By encompassing more extensive volumes, these larger simulations offer a broader and more representative view of the properties of the Universe. Although the present coverage of such simulations is still relatively limited, the ongoing improvement in the size of high-resolution simulations promises to enable more direct comparisons between observed and simulated galaxies. For example, recent work by \citet{Dubois2021} and \citet{Kim2022} has demonstrated that a more realistic treatment of observational selection effects through forward modelling of simulation data can alleviate some of the observed discrepancies between theoretical and observed galaxy stellar mass functions. In a similar vein, high resolution simulations in cosmological volumes are also able to resolve classical tensions between theory and observations at small physical scales \citep{Bullock2017} e.g. the existence of planarity in the kinematics of satellites around massive galaxies \citep{Ibata2013,Uzeirbegovic2024}. 




Cosmological simulations are typically calibrated to reproduce the properties of observed galaxy populations, but given their present observational incompleteness, this has not been possible in the low-mass regime. Dwarf galaxies, in contrast to massive galaxies, exhibit distinct evolutionary behaviour, dominated by stellar feedback and the influence of the local environment \citep{Watkins2023}. High-quality observations of the low-mass regime in combination with high-fidelity simulations can therefore provide important additional constraints on aspects of our galaxy evolution models such as stellar \citep{Dekel1986}, AGN \citep{Reines2013,Kaviraj2019,Davis2022}, and UV \citep{Haardt1996} feedback, which are expected to exhibit significantly higher efficiency in the shallower haloes of low-mass galaxies.

In anticipation of data from larger-scale studies, we use currently available deep data to present a preview of what will soon be possible on a much larger scale. Here we utilise ultra-deep HSC-SSP observations of the 2 square degree COSMOS field \citep{Scoville2007} in combination with two intermediate volume, high-resolution cosmological simulations -- \textsc{Illustris TNG50}  \citep{Nelson2019,Pillepich2019} and \textsc{NewHorizon} \citep{Dubois2021}. The focus of this paper is on measuring the morphologies of galaxies, which have been shown to correlate with their physical properties \citep[e.g.][]{Dressler1980,Dressler1997,Strateva2001,Hogg2002,Bundy2005,Conselice2006,Skibba2009,Bluck2014,Whitaker2015,Uzeirbegovic2020,Uzeirbegovic2022,Jang2023}, and thus probe underlying physical processes that drive their morphological evolution over cosmic time \citep{martin2018_sph}. This approach offers a means to constrain and enhance our understanding of these processes in the largely unexplored low-mass regime.

We begin in Section \ref{sec:data} with a detailed description of the observed and simulated data. Then, in Section \ref{sec:method}, we describe the techniques we employ to generate realistic HSC-SSP-like images from the simulation data, as well as our methodology for assessing galaxy structure and comparing visual similarity. Finally, in Sections \ref{sec:results} and \ref{sec:discussion}, we present a detailed comparison of the morphology of galaxies between the observed data and both simulations, discussing the implications that discrepancies between the simulations and observations may have for the accuracy of physical recipes implemented in each simulation.

Throughout this paper we adopt separate $\Lambda$CDM cosmologies consistent with the different values used by each simulation (see Section \ref{sec:sims}). For the observed COSMOS data we adopt a \citet[][]{Komatsu2011} cosmology with $\Omega_{\rm m}=0.272$, $\Omega_\Lambda=0.728$, $\Omega_{\rm b}=0.045$ and $H_0=70.4 \ \rm km\,s^{-1}\, Mpc^{-1}$.

\section{Data}
\label{sec:data}

In this paper, we present a comparative analysis of the structural properties of synthetic and observed dwarf galaxies. Our study incorporates realistic mock observations generated from the \textsc{NewHorizon} \citep{Dubois2021} and \textsc{Illustris TNG50} \citep{Nelson2019,Pillepich2019} simulations in addition to imaging and data products obtained from the COSMOS \citep{Scoville2007} and HSC-SSP \citep{Aihara2018} surveys. In the following sections, we provide an overview of each of these datasets.

\subsection{Simulations}
\label{sec:sims}

The theoretical component of our study makes use of two state-of-the-art cosmological simulations, \textsc{NewHorizon} and  \textsc{Illustris TNG50}. Despite simulating roughly similar volumes with similar fidelity, they diverge notably in their approaches to solving equations of (magneto)hydrodynamics and their sub-grid physical models.

Although both simulations implement models of black hole accretion and feedback, black hole growth is minimal for masses examined in this study in both simulations \citep{Dubois2015,Dubois2021,Voit2024,Peirani2024}. We therefore neglect the role of AGN from this overview and future discussion regardless of observational and theoretical evidence of the potential role of AGN the evolution of dwarf galaxies \citep[e.g.][]{Kaviraj2019,Koudmani2021,Davis2022}. 

Below, we give a concise summary of the key aspects of each simulation's approach.

\subsubsection{\textsc{NewHorizon}}

The \textsc{NewHorizon} simulation\footnote{\url{https://new.horizon-simulation.org/}} \citep[][]{Dubois2021} is a zoom-in of the (142~Mpc)$^{3}$ parent Horizon-AGN simulation \citep{Dubois2014,Kaviraj2017}. Within the original Horizon-AGN volume, a spherical volume with a diameter of 20~Mpc is defined, corresponding to a dark matter (DM) mass resolution and effective \textit{initial} gas mass resolution of $m_{DM}=1.2\times 10^6 \ \rm M_\odot$ and $m_{gas}=2\times 10^5 \ \rm M_\odot$. The stellar mass resolution is $1.3\times10^{4}{\rm M_{\odot}}$ and the smallest allowed spatial refinement (resolution) is $\sim34$~pc. Full details of the physics and galaxy formation model can be found in \citet{Dubois2021}.

\paragraph{Cosmology} \textsc{NewHorizon} adopts a $\Lambda$CDM cosmology consistent with \citet[][]{Komatsu2011} ($\Omega_{\rm m}=0.272$, $\Omega_\Lambda=0.728$, $\Omega_{\rm b}=0.045$, $H_0=70.4 \ \rm km\,s^{-1}\, Mpc^{-1}$).

\paragraph{Hydrodynamics} \textsc{NewHorizon} utilises the adaptive mesh refinement (AMR) code \textsc{Ramses} \citep[][]{Teyssier2002} and gas is evolved with a second-order Godunov scheme and the approximate Harten-Lax-Van Leer-Contact \citep[HLLC,][]{Toro1999} Riemann solver with linear interpolation of the cell-centred quantities at cell interfaces. 30--100 million leaf cells are used per level of refinement in the zoom-in region from level 12 to level 22 with the minimum physical size of cells kept approximately constant by adding an extra level of refinement at every doubling of the expansion factor.

\paragraph{Gas physics} Radiative cooling of primordial and metal-enriched gas occurs in the presence of a spatially uniform UV background beginning after redshift $z=10$, following \citep{Haardt1996}. Cooling is allowed down to $\sim 10^4\, \rm K$ through collisional ionization, excitation, recombination, Bremsstrahlung, and Compton cooling. Metal-enriched gas can cool further down to $0.1\, \rm K$ based on \citet{Sutherland1993} above $\ga 10^4\,\rm K$ and from \citet{Dalgarno72} below $\la 10^4\,\rm K$, allowing the simulation to partially resolve the multiphase nature of the ISM. The UV background is self-shielded in optically thick regions \citep[$n_{\rm H}\ga 0.01 \,\rm H\, cm^{-3}$;][]{Rosdahl2012}, with UV photo-heating rates reduced by a factor $\exp\left(-n_{\rm H}/n_{\rm shield}\right)$, where $n_{\rm shield}= 0.01 \,\rm H\, cm^{-3}$.

\paragraph{Star formation} Star formation proceeds above a density threshold of $n_{\rm H}> 10 \,\rm H\, cm^{-3}$ following a Schmidt law with a variable efficiency related to the cloud turbulent Mach number and virial parameter \citep{Kimm2017,Trebitsch2017,Trebitsch2020}. It favours the rapid formation of stars in dense, gravitationally collapsing medium with compressible turbulence and results in potentially higher instantaneous star formation rates and more bursty star formation histories.

\paragraph{Feedback from massive stars} Feedback from Type II supernovae (SNe) proceeds assuming all stars with stellar masses greater than $6~{\rm M_{\odot}}$ explode instantaneously after $5\,\rm Myr$. Each stellar particle is represented by a \citet{Chabrier2003} initial mass function (IMF) with each SN explosion releasing kinetic energy of $10^{51}\,{\rm erg}$. To account for the cumulative effect of clustered SN explosions on total radial momentum \citep{Thornton98}, the specific frequency of SNe is boosted by a factor of 2 to $\sim0.03 \, \rm M_\odot^{-1}$ \citep{Kim17,Gentry19}.

A mechanical SN feedback model \citep{Kimm2014,Kimm2015} is employed. In the Sedov-Taylor energy-conserving phase, the assumed specific energy is injected to the gas, since hydrodynamics will naturally capture the expansion of the SN and impart the correct amount of radial momentum. In the momentum-conserving phase radial momentum is imparted directly according to \citep{Thornton98}. This avoids artificially rapid radiative cooling caused by under-resolved cooling lengths, which would otherwise suppress the expansion of the SN bubble.

Other sources of stellar feedback such as stellar winds and Type Ia SNe are not implemented in the \textsc{NewHorizon} stellar feedback model.

\paragraph{Volume} The \textsc{NewHorizon} volume encompasses a single contiguous region of average density spanning approximately $\sim20$~Mpc. This choice strikes a balance between sampling diverse environments and achieving high resolution, although it does not include extremely dense or rarefied regions. As a result, the volume effectively probes field and group environments, while falling short of cluster environments. The maximum halo mass found in the \textsc{NewHorizon} volume is $M_{h}\sim10^{13}{\rm M_{\odot}}$.

\subsubsection{\textsc{Illustris TNG50}}

The Next Generation Illustris\footnote{\url{https://tng-project.org/}} (\textsc{IllustrisTNG}) is a suite of cosmological magnetohydrodynamical (MHD) simulations covering three different comoving volumes at varying resolutions. In this paper we use of the \textsc{TNG50} run \citep{Nelson2019,Pillepich2019} which simulates a volume with 50 comoving Mpc on a side, with a stellar and gas particle resolution of $8.5\times10^4~{\rm M_{\odot}}$ and a median spatial resolution for star-forming ISM gas of 100-140~pc. Full details of the physics and galaxy formation model can be found in \citet{Weinberger2017} and \citet{Pillepich2018}.

\paragraph{Cosmology} \textsc{TNG50} adopts a $\Lambda$CDM cosmology consistent with \citet{Planck2016} ($\Omega_{\rm m}=0.309$, $\Omega_\Lambda=0.691$, $\Omega_{\rm b}=0.049$, $H_0=67.7 \ \rm km\,s^{-1}\, Mpc^{-1}$).

\paragraph{(Magneto)hydrodynamics} \textsc{TNG50} uses the moving-mesh code \textsc{AREPO} \citep{Springel2010}. \textsc{AREPO} uses a finite volume method on an unstructured, moving, Voronoi mesh, with a directionally unsplit second order Godunov scheme \citep{Pakmor2016}. \textsc{TNG50} solves equations of idealised continuum MHD utilising cell-centred magnetic fields \citep{Pakmor2011} combined with an approximate HLLD Riemann solver \citep{Miyoshi2005}. 

\paragraph{Gas physics} Radiative cooling of primordial and metal-enriched gas occurs in the presence of a redshift-dependent, spatially uniform UV background, with corrections for self-shielding in the dense ISM \citep{Katz1992,Faucher2009}. Further redshift dependent cooling of metal-enriched gas from metal lines is allowed based on \cite{Smith2008,Wiersma2009}. \textsc{TNG50} does not directly resolve the multiphase structure of the ISM, but rather treats it using an idealised model \citep{Springel2003}.

\paragraph{Star formation} Star formation proceeds stochastically above a density threshold of $n_{\rm H}> 0.1 \,\rm H\, cm^{-3}$ with constant efficiency following the Kennicutt–Schmidt relation.

\paragraph{Feedback from massive stars} Stellar feedback follows the \citet{Springel2003} idealised multi-phase effective equation of state model, in which feedback energy from Type Ia and Type II SNe directly heats the ambient hot phase and returns metal enriched gas to the ambient ISM assuming a \citet{Chabrier2003} IMF. The model includes additional feedback from isotropically injected star formation driven kinetic winds including contributions from the asymptotic giant branch. The galactic-scale winds generated by stellar feedback are placed by hand and are decoupled from interacting hydrodynamically with the surrounding gas until they reach regions with densities significantly below the star formation threshold.

\paragraph{Volume} The \textsc{TNG50} volume models a smaller region at a higher fidelity compared to the rest of the \textsc{IllustrisTNG} suite. Spanning approximately $50$~Mpc, this region captures a relatively diverse range of environments, including voids and filaments, galaxy groups and poor clusters while excluding more extreme environments such as more massive clusters. The maximum halo mass found in the TNG50 volume is $M_{h}\sim10^{14}{\rm M_{\odot}}$.

\subsection{Observations}

For the observational component of our study, we make use of catalogues from the COSMOS survey as well as supplementary deep imaging from Hyper Suprime-Cam. The COSMOS field is extremely well covered by a wealth of photometric observations across the UV, optical and IR spectrum, allowing for precise determination of galaxy redshifts and physical properties.

Like \textsc{TNG50} and \textsc{NewHorizon}, COSMOS probes relatively average environments. For the redshift range $0.05<z<0.3$ and galaxy stellar masses $M_{\star}>10^{10.5}$~M$_{\odot}$, we calculate a galaxy number density of $0.0007$~Mpc$^{-3}$ for our COSMOS sample. This is smaller but still comparable to galaxy number densities obtained for larger fields at similar redshift ranges. For example, by integrating the galaxy stellar mass function for SDSS at $z<0.05$ presented in \citet{Baldry2008} we obtain a value of $0.0017$~Mpc$^{-3}$. For the VIMOS VLT Deep Survey (VVDS) mass function at $0.05<z<0.4$ presented in \citet{Pozzetti2007} we obtain a more similar value of 0.0008~Mpc$^{-3}$. In comparison, the number density of galaxies more massive than $10^{10.5}~{\rm M_{\odot}}$ at $z=0.25$ is 0.0039~Mpc$^{-3}$ and 0.0029~Mpc$^{-3}$ within the \textsc{NewHorizon} volume and TNG50 volume respectively, both slightly larger than the average number density obtained for SDSS.

\subsubsection{HSC-SSP}

We make use of \texttt{deepCoadd} $i$-band imaging from the third release of the Hyper Suprime-Cam Subaru Strategic Program \citep[HSC-SSP DR3; ][]{Aihara2022}\footnote{\url{https://hsc-release.mtk.nao.ac.jp/}}. Prior to the co-addition, a global third-order sky subtraction fit to 8k$\times8$k $(\sim20^{\prime})$ superpixels followed by a second sky subtraction using $256\times256$ $(\sim40^{\prime\prime})$ superpixels is performed for each exposure.

We cross-match positions from the COSMOS2020 catalogue using the \texttt{pdr3\_dud\_rev} catalogue. The DR3 catalogue uses \texttt{calexp} frames for the detection and segmentation of sources, these are generated by performing an additional local sky subtraction using $128\times128$ superpixels on the \texttt{deepCoadd} image. Source detection proceeds following the same maximum likelihood detection method used by SDSS and is described in detail in \citet{Bosch2018}. For imaging, we choose to use \texttt{deepCoadd} rather than \texttt{calexp} frames because these better preserve flux at scales larger than the local sky subtraction bin size. However, as discussed in the next section, this is not likely to be significant for our sample of galaxies.

In order to retain as complete a sample as possible, we restrict our study only to the central $1.5^{\circ}$ of the COSMOS field where imaging depth exceeds $\mu(10^{\prime \prime}\times10^{\prime \prime},3\sigma)=31$~mag~arcsec$^{-2}$ (see Section \ref{A:depth} for a more in-depth exploration).

\subsubsection{COSMOS2020}

We make use of data products from the COSMOS2020 CLASSIC catalogue \citep{Weaver2022}\footnote{\url{https://cosmos2020.calet.org/}}. COSMOS2020 source detections are performed using \textsc{SExtractor} \citep{Bertin1996} on a $\chi^{2}$ combined $izYJHK_{s}$ detection image \citep{Szalay1999} which incorporates deep $i$-band and $z$-band imaging from HSC-SSP DR2. The HSC-SSP $i$-band imaging provides exceptional depth compared to other imaging used in COSMOS2020, making it the primary factor in most source detections and therefore yielding very similar results to the \texttt{pdr3\_dud\_rev} catalogue.

The COSMOS2020 $\chi^{2}$ detection images are constructed from HSC-SSP DR2 \texttt{calexp} frames, which use the same global sky subtraction method as the DR3 \texttt{deepCoadd} frames. These frames do not include the additional $128\times128$ superpixel local sky subtraction employed in the DR3 \texttt{calexp} frames which are used to generate the DR3 catalogs. Due to improved depth and image quality, we choose to make use of DR3 \texttt{deepCoadd} frames rather than attempt to precisely match the COSMOS2020 catalogue by using identical DR2 \texttt{calexp} frames. Although the additional local sky subtraction step may result in some differences in the data used to construct the COSMOS2020 and HSC-SSP DR3 object catalogues, including over-subtraction of the faint outskirts of large objects, the reduction of the two datasets is largely consistent at the scale of objects considered in this work, which are smaller than the $128\times128$ bin size used for the DR3 local sky subtraction step.

In our analyses, we employ \textsc{LePhare} \citep{Arnouts1999,Ilbert2006,Arnouts2011} photometric redshift, stellar mass and rest-frame photometry estimates from the COSMOS2020 catalogue. COSMOS2020 utilises extensive coverage from photometric bands spanning UV, optical and IR data from GALEX, CFHT, Subaru, UltraVISTA and IRAC, which enables precise photometric redshift estimates that exceed a precision of 5 per cent, even for the faintest sources. Notably, for the brightest sources with $i<21$, COSMOS2020 achieves photometric redshift performance that approaches the accuracy of spectroscopic estimates, with a precision better than 1 per cent \citep{Weaver2022}.

\section{Method}
\label{sec:method}

\subsection{Imaging}

In the following sections, we detail the methodology employed to extract postage stamp images from the HSC-SSP survey and the generation of corresponding postage stamp images from the \textsc{NewHorizon} and TNG50 simulations.

\subsubsection{Observed galaxies}

We cut out $g$, $r$ and $i$-band postage stamps from the HSC-SSP \texttt{deepCoadd} data via the HSC-SSP DAS cutout tool. Cutouts are centred on the coordinates specified in the COSMOS2020 catalogue with an angular size of $1^{\prime}$.

\subsubsection{Synthetic galaxies}

For both simulations, we select objects from a single snapshot corresponding to the median redshift of our COSMOS2020 sample ($z\sim0.2$). While this does not take into account redshift evolution, the disparity in structural properties between the two simulations is considerably larger than the limited morphological changes that might occur between $z\sim0.25$ and $z\sim0.05$. Instead, to match the redshift distribution of the observed sample, we assign galaxies distances matching the redshift distribution of the observed sample. Our approach accounts for observational effects such as the reddening of the SED, cosmological dimming and evolution of angular size with redshift.

\textsc{NewHorizon} and \textsc{TNG50} catalogues utilise different structure finders, namely \textsc{(Adapta)HOP} \citep{Aubert2004} and \textsc{subfind} \citep{Springel2001}, respectively, for galaxy identification. To avoid selecting galaxy fragments we use the HOP rather than \textsc{AdaptaHOP} structure finder for \textsc{NewHorizon}, which has the effect of keeping all stellar substructures connected to the same main substructure. To ensure that we are relatively agnostic to the structure finder used and that extended substructures are preserved, images are generated using every star particle lying within a 100~kpc box centred around the centroid of each galaxy as defined by their respective structure finder.

\paragraph{Images} We follow almost the same procedure as \cite{Martin2022}, in this case to produce HSC-like, rather than Rubin-like, mock observations. We briefly outline our method below.

We begin by generating spectral energy distributions (SEDs) for each star particle extracted from within a 100~kpc box around the centroid of each galaxy using a grid of \citet[][]{Bruzual2003} simple stellar population (SSP) models interpolated to the age and metallicity of each star particle and assuming a single \citet[][]{Chabrier2003} IMF. Dust attenuation is calculated for the SED of each stellar particle via a screen model which depends on the dust column density summed along the line of sight to each star particle, calculated assuming a gas-to-dust ratio of 0.4 \citep[e.g.][]{Draine2007}. An $R = 3.1$ Milky Way dust grain model \citep[][]{Weingartner2001} is then used to produce the final dust attenuated SED for each particle. As \citet{Watkins2025} show, low gas column densities mean that the precise dust model adopted has very limited influence on the broadband photometry of galaxies in the mass range considered here. Adopting an SMC extinction curve and varying the dust-to-metal ratios between 0.01 and 1.0 introduces a typical variation in the central surface brightness of the order $0.01$~mag arcsec$^{-2}$.

This approach accounts for the geometry of the spatial distribution of dust within and around the galaxy, but does not account for other effects such as scattering. However, our exact treatment of dust is unlikely to have any significant impact on galaxy structure, especially, at the low redshifts probed by this study, where gas-rich starbursts are rare \citep[e.g.][]{Atek2014}. For the mass ranges considered in this study, we find that a majority of galaxies have very low dust column densities, which result in 95 per cent of galaxies having $i$-band magnitudes attenuated by less than 0.2 mag, with a majority of these being attenuated by a completely negligible amount. Additionally, due to the redshift distribution of our sample, for which more than $90$ per cent of galaxies are found at redshifts $z>0.1$, structures produced by dust are not well resolved in the vast majority of cases. Dust lanes are only expected to begin to appear at stellar masses of $10^{9}~{\rm M_{\odot}}$ \citep{Holwerda2019}, which is close to the end of the mass range considered in this study. Furthermore visual inspection of a subset of observed images from HSC-SSP in the mass and redshift range covered by our sample, does not yield any clear dust structures.

We finally create a 2-d image by first redshifting and convolving each SED with the HSC $i$-band transmission functions \citep{Kawanomoto2018} and then adaptively smoothing the particles in 3D to better represent the distribution of stellar mass in phase space and remove unrealistic variation between adjacent pixels due to resolution effects. The smoothed distribution is summed along one of the axes and binned into $0.168^{\prime\prime}$ pixels to produce 2-d $g$, $r$ and $i$-band flux maps, which are then convolved with the HSC PSF \citep{Montes2021} and scaled so that the magnitude is given by $i = -2.5{\rm log_{10}}(f)$ where $f$ is the flux in each pixel.


\paragraph{Source injection}

To enhance the realism of our synthetic images, we then inject them into real backgrounds drawn from the HSC-SSP data. This ensures that the distribution of depths in our synthetic data closely matches that of the observed data as well as incorporating a realistic distribution of nearby background objects, which could potentially introduce scatter or systematic biases into our measurements.

\begin{figure}
    \centering
    \includegraphics[width=0.45\textwidth]{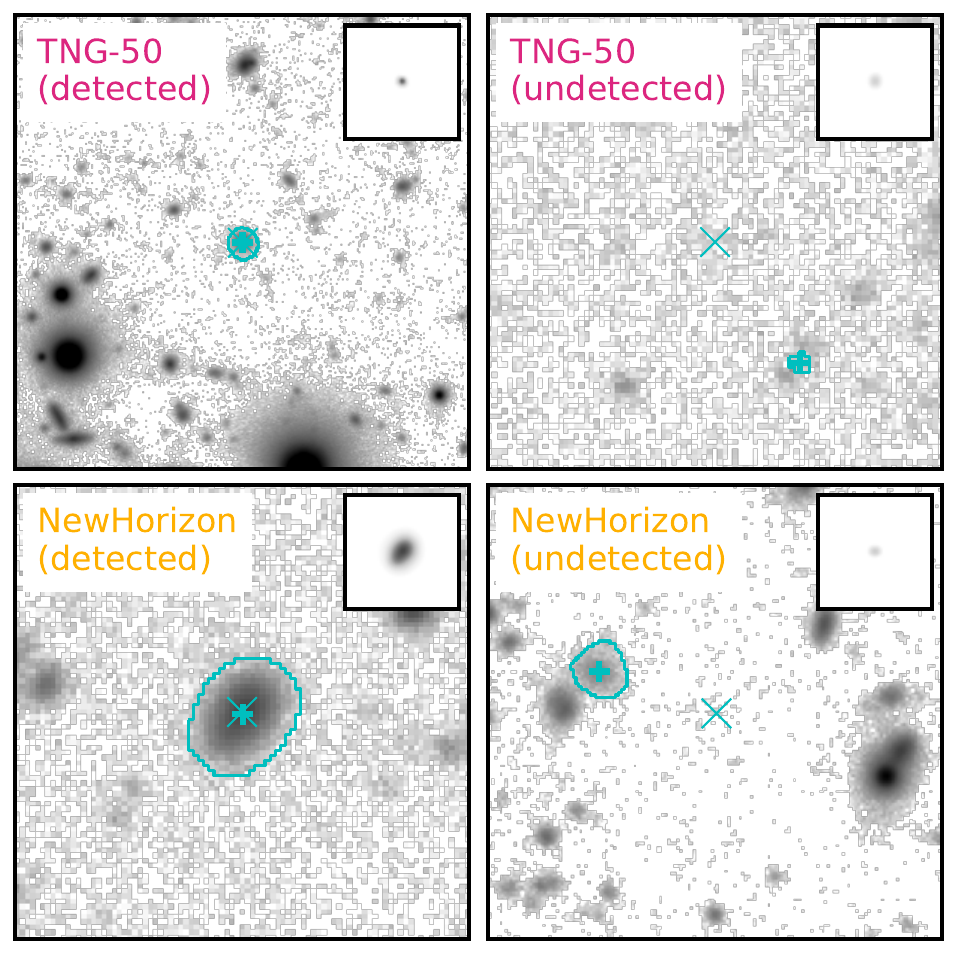}
    \caption{Examples of synthetic galaxy injections onto HSC-SSP \texttt{deepCoadd} data. The centroid of the injected galaxy is indicated with a cyan 'X' while thicker cyan crosses indicate the centroid of the nearest detected object, with cyan contours showing the segmentation map for this object. The inset panel shows the injected object. All images have the same physical scale at the redshift of the injected galaxy, therefore the angular scale differs between images.}
    \label{fig:injection}
\end{figure}

In the central region of the COSMOS field, we identify suitable sites for injecting synthetic sources. Using the Astropy package \textsc{PhotUtils} \citep{Bradley2022}, we select locations that are sufficiently distant from any detected pixels. To do this, we first estimate the 2-d background using a median background estimator with $3\sigma$ clipping and then apply a detection threshold of 1.5 times the background RMS ($1.5\sigma$). This approach minimizes the risk of chance projections between real sources and our injected synthetic sources.

We then generate a detection map for multiple randomly selected regions within the central COSMOS HSC-SSP \texttt{deepCoadd} images. Each image is first convolved with a Gaussian kernel with a standard deviation matching the measured FWHM. We then apply the $1.5\sigma$ threshold to identify candidate injection sites, ensuring they are at least 20 pixels away from any detected pixels.

Each selected point serves as the centroid for injecting synthetic galaxies. Since the initial regions are randomly selected, we disregard any effects from correlated structure in the simulations.

The HSC-SSP \texttt{deepCoadd} images are scaled to a zero-point of 27 mag. After convolution, we rescale our synthetic images using the same \texttt{fluxmag0} parameter to ensure their fluxes match the image DNs of the HSC data (in units of nJy) before injecting them at the designated centroids.

Since our procedure involves injecting objects into pre-processed \texttt{deepCoadds}, the impact of the injected objects on sky subtraction is not accounted for. However, given their faintness and small sizes relative to the superpixels used for background modeling, we do not expect a significant effect on either sky subtraction or detectability.

\subsection{Detection \& segmentation of synthetic \& observed galaxies}
\label{sec:segmentation}

After injecting each synthetic object, we use \textsc{PhotUtils} to generate a new detection map following the same procedure described above. Once the detection map is created, we deblend it to produce a segmentation map using the \textsc{PhotUtils} \texttt{deblend\_sources} procedure which utilises a combination of multi-thresholding and watershed segmentation. For source deblending, we set a minimum galaxy area of 10 pixels, use 32 deblending levels, and adopt a minimum contrast of 0.001.

For each object identified in the segmentation map, we determine its centroid and compare it to the centroid of the injected galaxy. A synthetic object is considered detected if the segmentation map contains an object with a centroid located within either 10 pixels or one effective radius of the injection centroid. Since our selection criteria require no detected pixels within a 20-pixel radius of the injection centroid, the likelihood of a detection resulting from a chance projection is significantly reduced.

Figure \ref{fig:injection} illustrates examples of both detections and non-detections for \textsc{TNG50} and \textsc{NewHorizon} galaxies following their injection into HSC-SSP $i$-band data.

Rather than relying on the footprints provided by the HSC-SSP data release, we follow an almost identical procedure for detecting and segmenting observed galaxies from the HSC-SSP \texttt{deepCoadd} images by again generating a detection map and performing source deblending using \textsc{photutils}. To ensure that the object segmentation maps we generate using \textsc{photutils} correspond well to the object footprints used to calculate the quantities in the COSMOS2020 catalogue, we require that the centroids from the COSMOS2020 catalogue match the centroids of the recovered sources to within 10 pixels.

\subsection{Sample selection}

We first select a sample of dwarf galaxies using the \textsc{LePhare} derived photometric redshifts and masses taken from the COSMOS2020 catalogue. Galaxies are selected only from the region of COSMOS with the deepest HSC $i$-band imaging (better than 31~mag~arcsec$^{-2}$, Appendix \ref{A:depth}), and with the following properties:
\begin{enumerate}
    \item photometric redshift in the range $0.05 < z < 0.25$
    \item fractional redshift error smaller than 10 per cent
    \item $10^{7.5}$ < $M_{\star}$/$\rm{M_{\odot}}< 10^{9.5}$
    \item Within the central $1.5^{\circ}$ of the COSMOS field.
\end{enumerate}
\noindent yielding a sample of 1320 galaxies.

\begin{figure}
    \centering
    \includegraphics[width=0.45\textwidth]{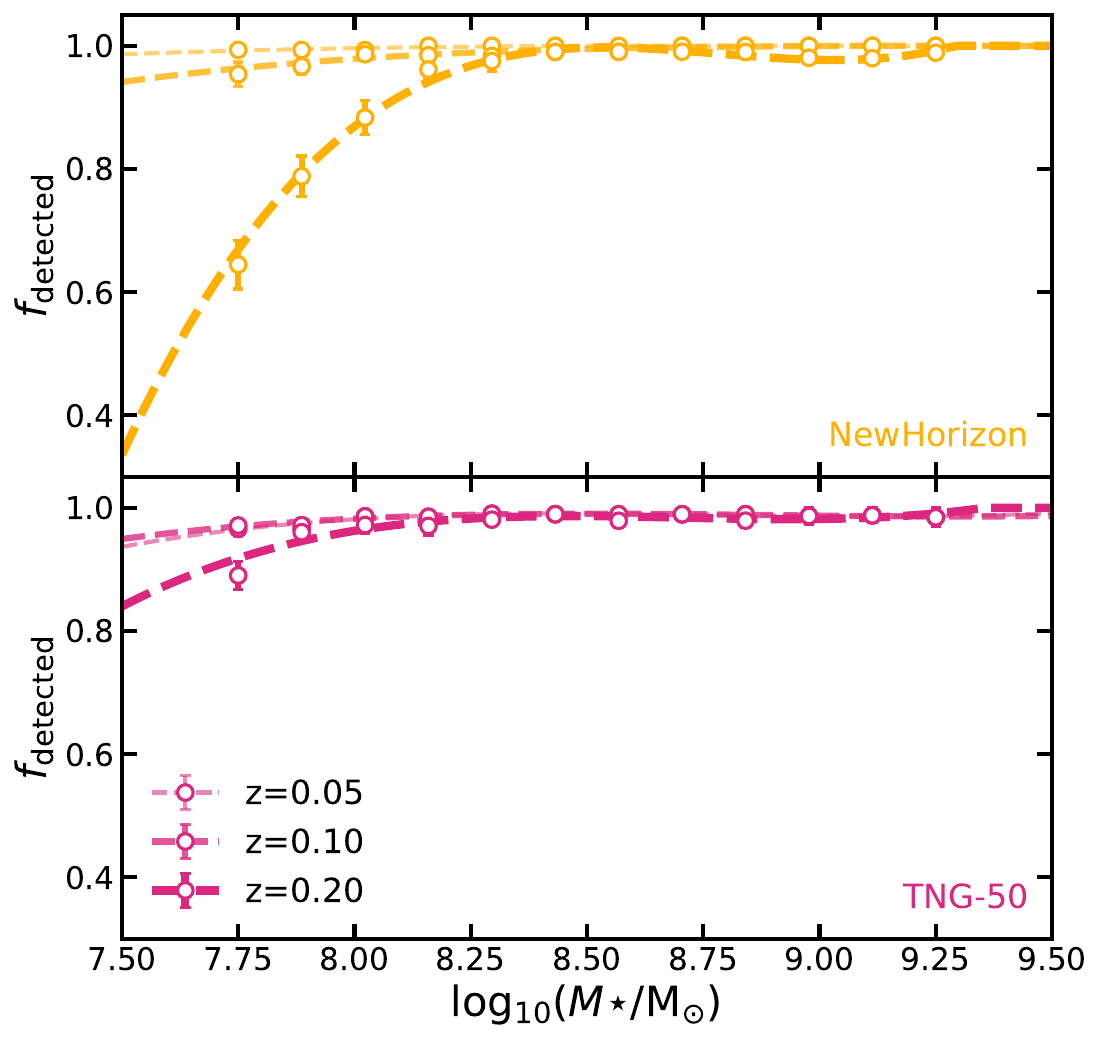}
    \caption{Fraction of galaxies with successful detections as a function of stellar mass following injection of the same sample of synthetic galaxies into HSC-SSP \texttt{deepCoadd} at three different redshifts indicated by the legend. Dashed lines show polynomial fits to the binned points. The top panel shows the results for \textsc{NewHorizon} and the bottom panel shows the results for \textsc{TNG50}.}
    \label{fig:detected}
\end{figure}

To evaluate the expected observational completeness of our sample. we conduct an analysis of detection rate of galaxies from each simulation injected into the HSC-SSP COSMOS field. Figure \ref{fig:detected} shows the fraction of detected galaxies as a function of stellar mass and redshift for both the \textsc{NewHorizon} and \textsc{TNG50} simulations. In the case of both simulations, we find that nearly 100 percent of galaxies are detected across the entire mass range of our sample up to a redshift of $z=0.1$. However, it is worth noting that \textsc{NewHorizon} galaxies exhibit a noticeable decrease in completeness at higher redshifts (which is likely a result of their more diffuse structure as shown in Section \ref{sec:results_sersic}). Specifically, at $z>0.2$, \textsc{NewHorizon} galaxies begin to exhibit incompleteness for stellar masses below $10^{8}{\rm M_{\odot}}$, with completeness levels dropping to around 40 percent for galaxies with a stellar mass of $10^{7.5}{\rm M_{\odot}}$.

Within the central region of the COSMOS field and within the mass and redshift ranges of our sample, completeness is expected to be high. However, at higher redshifts, we begin to observe some incompleteness. Notably, at low redshifts ($z<0.1$), both simulations demonstrate exceptionally high completeness, suggesting that neither simulation predicts the existence of a significant population of galaxies that would remain undetected when accounting for selection effects. As we show in Section \ref{sec:results}, the average S\'ersic indices and effective radii of COSMOS galaxies fall between the ranges defined by \textsc{NewHorizon} and \textsc{TNG50}. Consequently, we anticipate the completeness of COSMOS galaxies to lie somewhere between the completeness levels seen in the two simulations \citep[see e.g.][for discussion of the completeness of the HSC-SSP \textit{wide} layer as a function of S\'ersic parameters]{Greco2018_thesis,Greene2022}.

\begin{figure}
    \centering
    \includegraphics[width=0.45\textwidth]{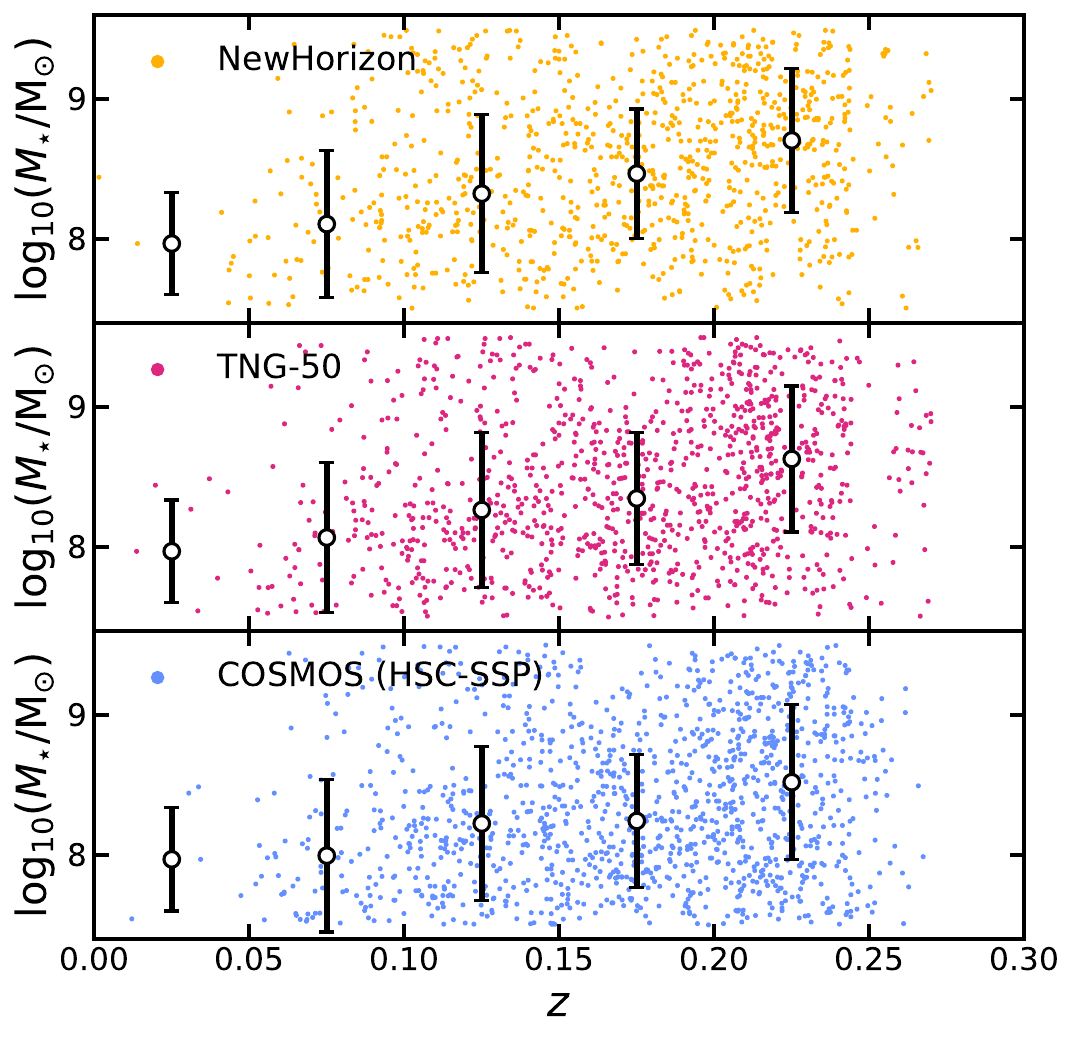}
    \caption{The stellar mass vs redshift distributions of the matched \textsc{NewHorizon} (top panel) and \textsc{TNG50} (middle panel) distributions compared with the original COSMOS sample (bottom panel). Black error bars with white circles show the $1\sigma$ dispersion and median of each distribution for bins of redshft,}
    \label{fig:z_dist}
\end{figure}

To replicate the observational biases present in the HSC-SSP data, we generate a redshift and mass matched sample of objects for both the \textsc{NewHorizon} and \textsc{TNG50} simulations. We allow simulated galaxies to be drawn more than once, provided they are observed at a different redshift and in a different orientation. Both synthetic samples contain the same number of objects as the COSMOS sample after any undetected objects are rejected. Figure \ref{fig:z_dist} shows the mass and redshift distribution of these matched samples.

In order to understand any biases in our measurements, we additionally create `rest-frame' images for the same mass-matched sample of \textsc{NewHorizon} and \textsc{TNG50} galaxies in which the pixel scale of the images are set to a fixed physical size of $1~{\rm kpc/{\prime\prime}}$, equivalent to 0.168~kpc per pixel given the HSC pixel size. Stellar SEDs are also kept in the rest-frame but the the images are otherwise produced and processed identically.

\subsection{Parametric and non-parametric galaxy structure measures }

In this study, we employ non-parametric Gini, $M_{20}$, and CAS measures of galaxy structure, as well as single-component S\'ersic fits, to analyse both simulated and observed galaxies \citep{Conselice2003}.

To calculate these parameters, we make use of the \textsc{statmorph} Python package \citep{Rodriguez2019}. \textsc{statmorph} is a tool designed for the analysis and characterisation of structural properties of galaxies and has been previously employed to investigate both simulated and observed galaxies \citep[e.g.][]{Rodriguez2019,Sazonova2020,Kartaltepe2023,Ortega2023}. Simulation-based studies \citep{Rodriguez2019,Ortega2023} have focused on massive galaxies at low redshift ($z<0.05$) using both the \textsc{Illustris TNG50} simulation and the larger volume, lower resolution TNG100 simulation. We adopt a slightly different approach to these studies in our analysis. As described earlier, we insert galaxies into observed backgrounds and then segment and perform measurements using the same methodology applied to the observed galaxies. This allows us to take into account observational biases, exclusively measuring the properties of galaxies that would be detectable in observational data.

Non-parametric measures are sensitive to imaging resolution and depth \citep[e.g.][]{Sazonova2024}, making it challenging to compare the properties of objects observed at different redshifts. By selecting a sample with matched redshift and mass distributions, we mitigate this issue. However, individual measurements will still exhibit redshift dependence, as demonstrated by the rest-frame quantities presented in Appendix \ref{A:restframe}.

In the subsequent sections, we provide a concise overview of the measures employed throughout the remainder of this paper. We discard all galaxies where the S\'ersic index exceeds $n=4$. As well as being rare in the nearby Universe in the low-mass regime \citep{Su2021,Seo2022,Watkins2022,Lazar2024a}, many small, high S\'ersic index objects will appear as point sources at the redshift and spatial resolution employed in this study and therefore likely to be poor fits.

\subsubsection{S\'ersic profile fitting}

The S\'ersic profile \citep{Sersic1968} parameterises the light profile of a galaxy. It encapsulates several important characteristics of the light profile, including the effective radius ($R_{\rm eff}$), the surface brightness at the effective radius ($\mu_{\rm eff}$), the ellipticity ($e$) and the S\'ersic index ($n$), which parameterises the central concentration of the profile.

To properly account for the impact of seeing, we take advantage of \textsc{statmorph}'s ability to convolve the S\'ersic model with a PSF while fitting. In our study, we use the same HSC $i$-band PSF obtained from \citet{Montes2021} which we previously used to generate our synthetic observations. Because there is some variation in the PSF across the HSC-SSP COSMOS field, we rescale the PSF model when fitting to our observed sample. For each object, we rescale the \citet{Montes2021} PSF to match the FWHM measured at the centre of the specific patch where each galaxy is located, as indicated in the \texttt{iseeing} column of the \texttt{patch\_qa} table within the HSC-SSP catalogue.

\subsubsection{Gini-$M_{20}$}
The Gini-$M_{20}$ classification system \citep{Lotz2004} is widely used in astronomy both to identify mergers or irregular galaxies and as a general measure of galaxy morphology.

\paragraph{Gini}
The Gini statistic \citep{Glasser1962} quantifies the homogeneity of the light distribution within a galaxy by comparing its sorted cumulative distribution function to the expectation of an even flux distribution. In the context of this study, the Gini statistic is adapted from the original definition by \citet{Glasser1962} to account for negative flux values, requiring a first-order correction \citep{Lotz2004}. Lower Gini values correspond to galaxies with more uniform light distributions, such as late-type spirals with extended disks, while higher values indicate centrally concentrated systems, such as early-type galaxies or compact starbursts.

\paragraph{$M_{20}$}
The $M_{20}$ statistic \citep{Lotz2004} measures the spatial variance of the brightest regions in a galaxy relative to the entire galaxy. It compares the intensity-weighted central second moment of the galaxy with the sum of the second moments of the pixels that containing the brightest 20 percent of the total galaxy flux. Lower values of $M_{20}$ correspond to more centrally concentrated galaxies, such as ellipticals or early-type spirals, where the brightest regions are closer to the center. Higher values of $M_{20}$ indicate more spatially extended bright regions, as seen in late-type spirals and irregular galaxies, often associated with star formation clumps or disturbed morphologies.

\subsubsection{Concentration, Asymmetry and Smoothness (CAS)}
\label{sec:CAS}

Another commonly used non-parametric method for quantifying galaxy structure is based on the CAS (Concentration, Asymmetry, and Smoothness) statistics \citep{Conselice2003}. Our analysis is conducted within \textsc{statmorph}'s default circular aperture size set at 1.5 Petrosian radii ($R_{\rm P}$).

\paragraph{Concentration}
The concentration parameter compares the radii enclosing 80 per cent and 20 per cent of a galaxy’s total flux, reflecting the steepness of its surface brightness profile. Disk-like galaxies tend to have lower concentrations, while bulge-dominated and early-type systems have higher values. Mergers can span a wide range, with early-stage interactions appearing more diffuse and late-stage mergers becoming more centrally concentrated.

\paragraph{Asymmetry}
The asymmetry parameter quantifies deviations from symmetry by comparing a galaxy’s flux distribution to its 180 degree rotated counterpart, with a background correction applied. Regular elliptical and disk galaxies tend to be more symmetric, while interacting or merging systems, irregular galaxies, and those with prominent star-forming regions or dust lanes exhibit higher asymmetry.

\paragraph{Smoothness}
The smoothness parameter measures the contribution of small-scale structures to a galaxy’s light distribution. Smooth, well-ordered systems like ellipticals have lower smoothness, while clumpy or irregular galaxies, including mergers and star-forming systems, show higher values. However, due to the small angular sizes of most galaxies in our sample, we do not use smoothness, as reliable estimates are difficult to obtain.

In our final samples, we discard any objects flagged as having poor quality fits for either the S\'ersic measurements or any of the other morphological measurements. We observe a comparable reduction in sample size in each dataset: a decrease of 12 per cent, 12 per cent, and 8 per cent for \textsc{NewHorizon}, \textsc{TNG50}, and COSMOS, respectively. This similar reduction in both simulated and observed data suggests that there is no notable bias in terms of the quality of fits between them.

\section{Results}
\label{sec:results}

In this section, we present a comparison of the structural and visual properties of our matched synthetic and observed dwarf galaxy samples. We note that this study consists of a comparison of the statistical projected properties of galaxies across a relatively large redshift range which introduces variations in individual measurements due to factors such as seeing, redshifted galaxy SEDs, cosmological dimming, and evolution of the angular size distance. This is somewhat mitigated in the case of the S\'ersic fits as the model is convolved with the PSF, but no corrections are applied to other quantities.

In Appendix \ref{A:restframe}, we present rest-frame measurements of the simulated galaxies at a fixed angular scale. This facilitates more direct comparisons between the two simulations, which we also compare with a sample of COSMOS galaxies at $z<0.1$. We note that, while we observe consistent qualitative trends between the observed- and rest-frame quantities of our S\'ersic fits, there are some differences in some of the other quantities, driven primarily by the very compact sizes of some TNG50 galaxies. These differences and their causes are discussed in detail throughout this section.

\subsection{S\'ersic profile fitting}

\label{sec:results_sersic}

In this section we compare the structural properties of simulated and observed galaxies estimated from single component S\'ersic fits. 

\begin{figure}
    \centering
    \includegraphics[width=0.45\textwidth]{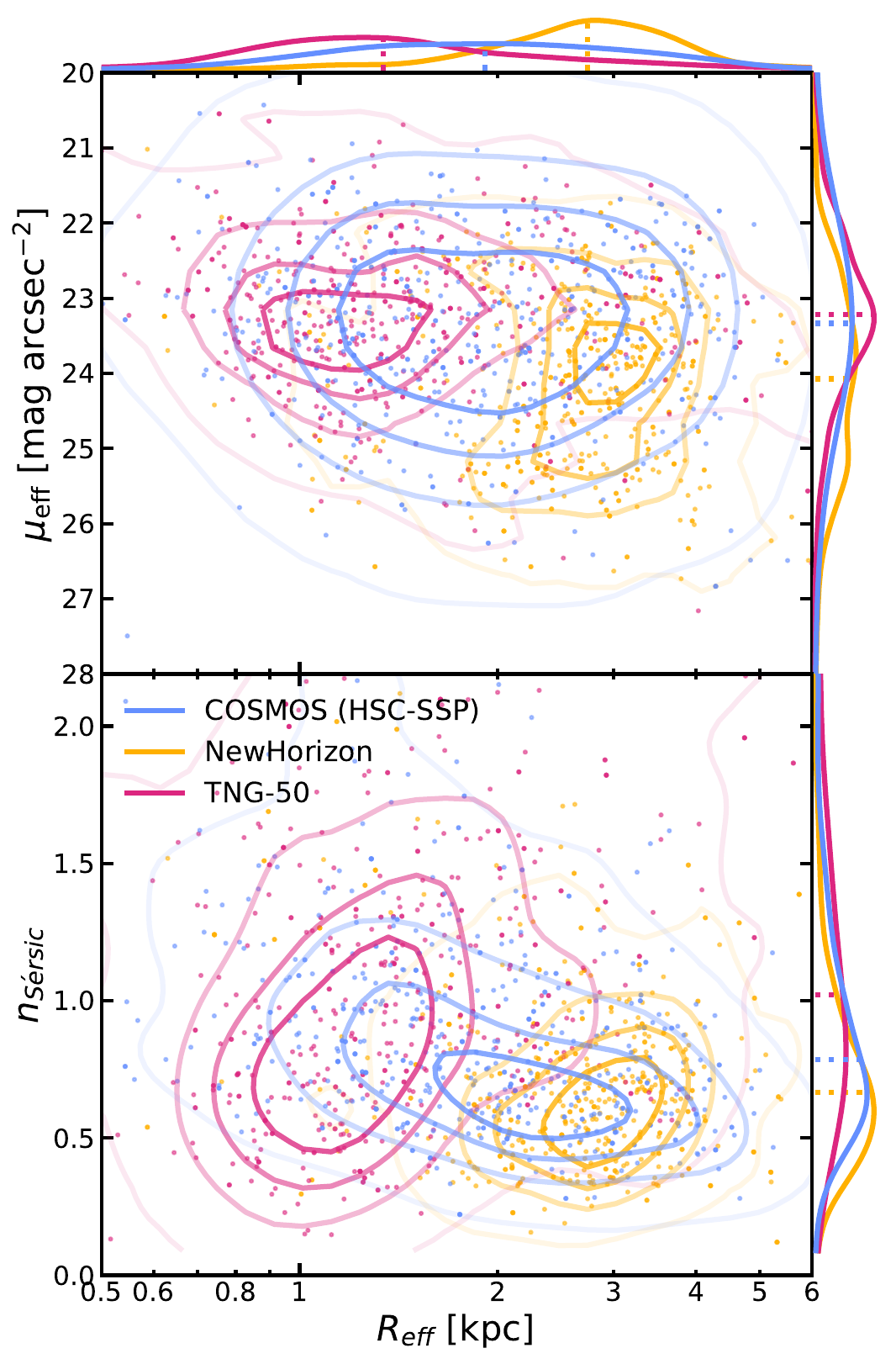}
    \caption{Contour plots showing the 2-d distribution of surface brightness at the effective radius (top panel) and S\'ersic index (bottom panel) as a function of effective radius for COSMOS (blue) with the redshift and mass-matched samples from \textsc{NewHorizon} (yellow) and \textsc{TNG50} (red). Coloured points show a randomly selected sub-sample with the same colour scheme. The sides of each panel show the marginal distribution of each variable with dashed lines indicating the median values. See Figure \ref{fig:sersic_rest} for a plot showing the rest-frame quantities at fixed angular scale.}
    \label{fig:sersic}
\end{figure}

Figure \ref{fig:sersic} shows the 2-d distribution of S\'ersic parameters for COSMOS (blue) and the redshift and mass-matched samples from \textsc{NewHorizon} (yellow) and \textsc{TNG50} (red). The top and bottom panels indicate the distribution of surface brightness at the effective radius ($\mu_{\rm eff}$) and S\'ersic index ($n_{\rm S\acute{e}rsic}$) respectively as a function of effective radius ($R_{\rm eff}$) in the form of scatter and contour plots. The marginal distribution of each parameter is shown along the sides of each panel, with dashed lines indicating the median value of each distribution.

Except for the case of $\mu_{\rm eff}$, average values for \textsc{NewHorizon} and \textsc{TNG50} galaxies lie at opposite extremes of the COSMOS distribution. \textsc{NewHorizon} dwarfs exhibit significantly larger sizes, evident from high $R_{\rm eff}$ and $n_{\rm S\acute{e}rsic}$ values, while \textsc{TNG50} dwarfs appear much more compact. At higher masses than those probed in this study, the size-mass relation of galaxies for both simulations has been shown to closely match observations \citep{Dubois2021,Wang2023}, indicating these differences are specific to the low-mass regime.

In contrast to COSMOS and \textsc{TNG50}, \textsc{NewHorizon} also shows a skewed distribution in $\mu_{\rm eff}$ that extends to low surface brightnesses. COSMOS galaxies exhibit brighter $\mu_{\rm eff}$ on average than both \textsc{NewHorizon} and \textsc{TNG50} galaxies, although, as we discuss later, this is likely to be quite sensitive to both the assumptions made when producing the mock images and the chemical and star-forming properties of the galaxies. But given the generally high completeness of each sample, it is unlikely that there is a strong effect on the distribution of any of the other S\'ersic parameters.

\begin{figure}
    \centering
    \includegraphics[width=0.45\textwidth]{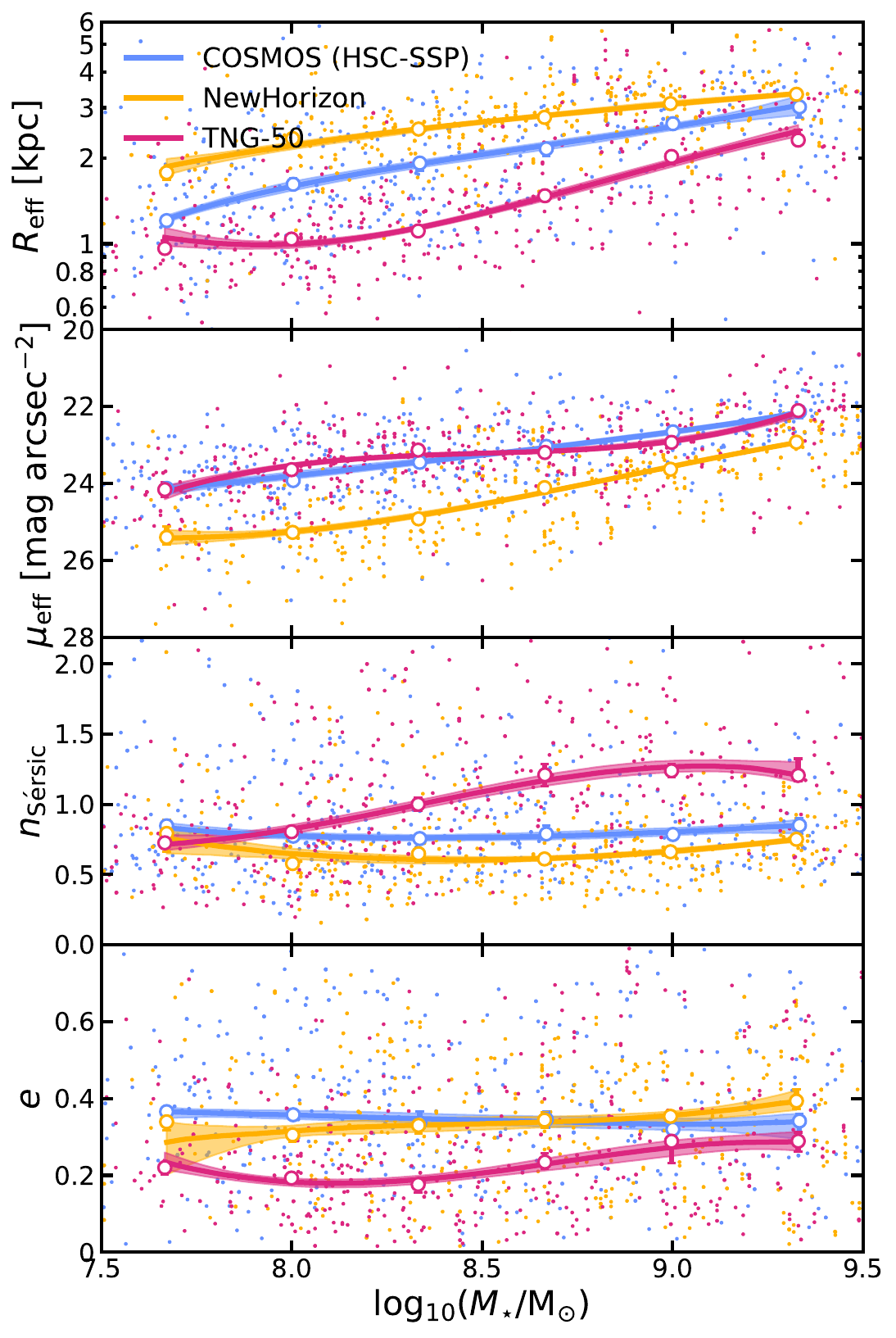}
    \caption{Plots showing the trend in the median effective radius, surface brightness at the effective radius, S\'ersic index and projected ellipticity as a function of stellar mass for COSMOS (blue) and the redshift and mass-matched samples from \textsc{NewHorizon} (yellow) and \textsc{TNG50} (red). Open circles with error bars show the median and error on the median for individual redshift bins, with filled regions indicating the $1\sigma$ uncertainty. Coloured points show a randomly selected sub-sample with the same colour scheme. See Figure \ref{fig:rest_sersic_mass} for a plot showing the rest-frame quantities at fixed angular scale.}
    \label{fig:sersic_mass}
\end{figure}

Figure \ref{fig:sersic_mass} shows the trend in the median value of each S\'ersic parameter with stellar mass, with the addition of the projected ellipticity ($e$). For each parameter \textsc{NewHorizon} and COSMOS galaxies both exhibit similar trends with stellar mass (albeit with different normalisation). \textsc{TNG50} show strong trends in $R_{\rm eff}$, $n_{\rm S\acute{e}rsic}$ and ellipticity. Galaxies appear significantly rounder at lower masses, but rise to similar values to those of \textsc{NewHorizon} and COSMOS at the high-mass end. In the rest-frame (Figure \ref{fig:rest_sersic_mass}), there is a less significant difference, likely arising from the fact that \textsc{NewHorizon} galaxies are physically larger so less susceptible to being smeared by the PSF. The same does not appear to be true for $n_{\rm S\acute{e}rsic}$, which is still seen to evolve quite significantly both in the mass-matched sample and in the rest-frame. Finally, both the mass-matched and rest-frame $R_{\rm eff}$ for \textsc{TNG50} begins to plateau at low mass $(M_{\star}\lesssim10^{8}~{\rm{M_{\odot}}})$. We note that, given the coarser resolution of \textsc{TNG50} compared with \textsc{NewHorizon} (100-140~pc vs 30-40~pc) galaxies are sampled by relatively small number of resolution elements below an effective radius of 1~kpc.

\subsection{Gini and $M_{20}$}

\label{sec:results_gini}

We also compare the non-parametric Gini and $M_{20}$ \citep{Lotz2004} measures. These metrics evaluate the homogeneity of the light distribution and the concentration of the brightest regions in relation to the centre of the light distribution, respectively.

\begin{figure}
    \centering
    \includegraphics[width=0.45\textwidth]{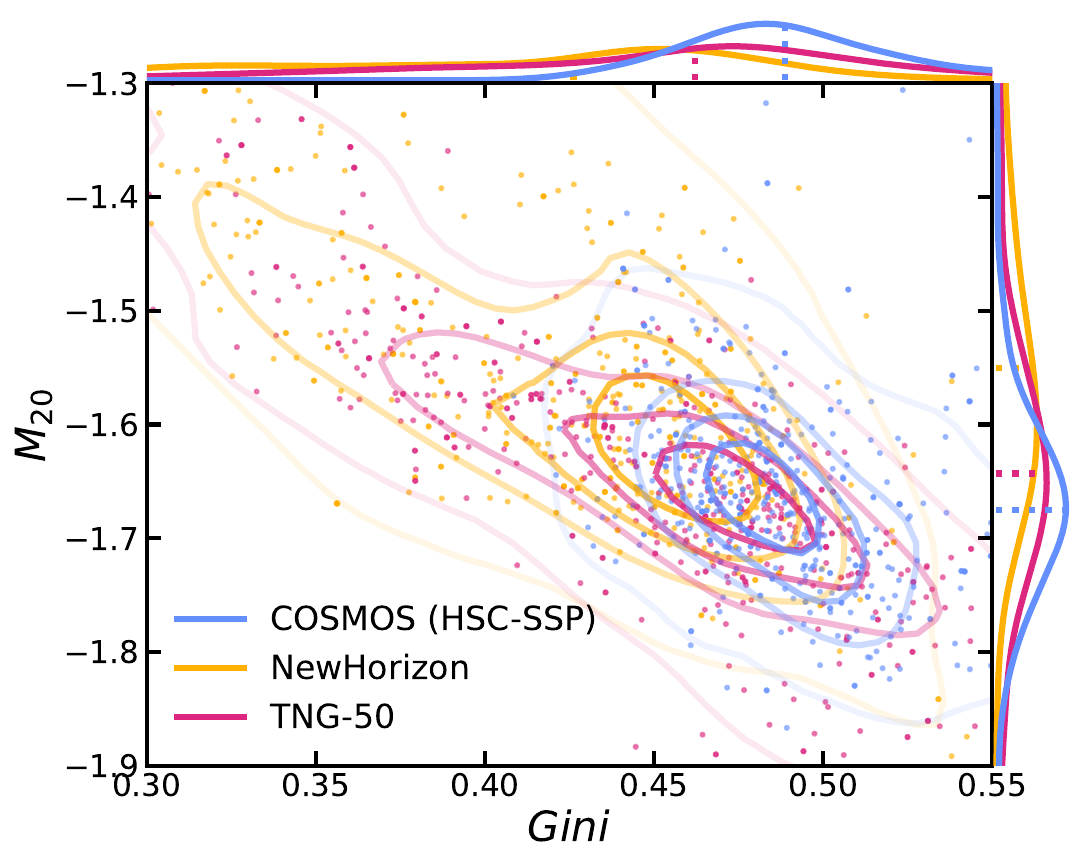}
    \caption{Contour plot showing the 2-d distribution of Gini and $M_{20}$ for COSMOS (blue) and the redshift and mass-matched samples from \textsc{NewHorizon} (yellow) and \textsc{TNG50} (red). Coloured points show a randomly selected sub-sample with the same colour scheme. The sides of the plot show the marginal distribution of each variable with dashed lines indicating the median values. See Figure \ref{fig:giniM20_rest} for a plot showing the rest-frame quantities at fixed angular scale.}
    \label{fig:giniM20}
\end{figure}

Figure \ref{fig:giniM20} shows the 2-d distribution of Gsini and $M_{20}$ for COSMOS (blue) and the redshift and mass-matched samples from \textsc{NewHorizon} (yellow) and \textsc{TNG50} (red). The marginal distribution of each parameter is along the edges of the plot, with dashed lines indicating the median value of each distribution.

\textsc{NewHorizon} and \textsc{TNG50} share similar distributions although \textsc{NewHorizon} galaxies are considerably more skewed towards high Gini values and high $M_{20}$ values indicating more uneven light distributions. COSMOS dwarfs, on the other hand, exhibit a narrow range of values concentrated towards higher Gini values and lower $M_{20}$ values, indicating smoother, more centrally concentrated light distributions.

However Figure \ref{fig:giniM20_rest} reveals that, in the rest-frame, TNG50 dwarfs are shifted towards higher Gini values and lower $M_{20}$ due to their intrinsically smaller sizes, which result in their more tightly concentrated central regions being more effectively smeared out by the PSF. Overall, TNG50 dwarfs are actually considerably more centrally concentrated than COSMOS galaxies, even compared with the COSMOS $z<0.1$ sample. This better agrees with the results of our S\'ersic fits which include a correction for the PSF and show that \textsc{TNG50} galaxies have steeper S\'ersic indices than COSMOS galaxies.

\begin{figure}
    \centering
    \includegraphics[width=0.45\textwidth]{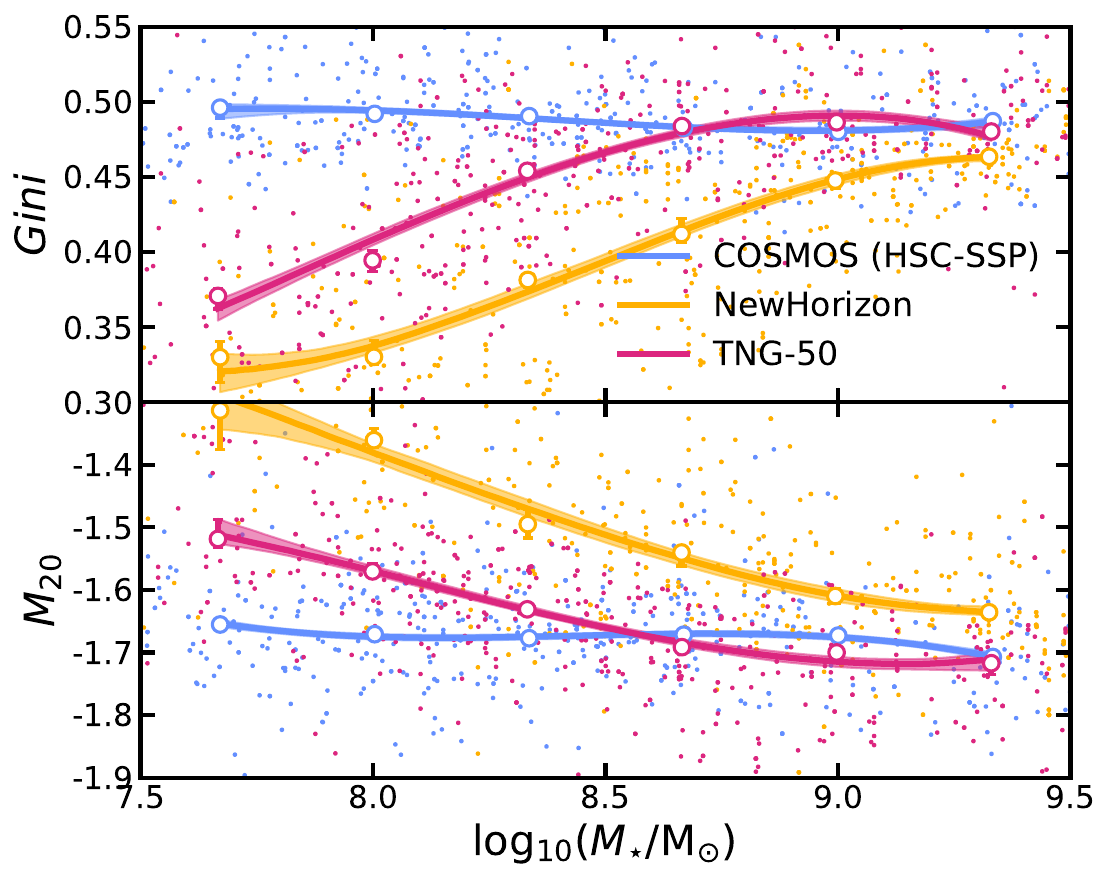}
    \caption{Plots showing the trend in the median Gini and $M_{20}$ as a function of stellar mass for COSMOS (blue) and the redshift and mass-matched samples from \textsc{NewHorizon} (yellow) and \textsc{TNG50} (red). Open circles with error bars show the median and error on the median for individual redshift bins, with filled regions indicating the $1\sigma$ uncertainty. Coloured points show a randomly selected sub-sample with the same colour scheme. See Figure \ref{fig:rest_ginim20_mass} for a plot showing the rest-frame quantities at fixed angular scale.}
    \label{fig:giniM20_mass}
\end{figure}

Examining the trend with mass in Fig. \ref{fig:giniM20_mass}, COSMOS dwarfs show no notable change across the mass range, maintaining consistently high Gini values and low $M_{20}$ values. Both \textsc{TNG50} and \textsc{NewHorizon} galaxies display similar trends with stellar mass, with higher Gini and lower $M_{20}$ values toward higher mass. This trend is qualitatively replicated in the rest-frame, except for the $M_{20}$ of \textsc{NewHorizon}, which has flat trend with stellar mass.

\subsection{Concentration and Asymmetry}

\label{sec:results_CAS}

Finally we compare the concentration and asymmetry parameters from the CAS \citep{Conselice2003} system for simulated and observed galaxies. Concentration and asymmetry serve as metrics to assess the degree of central concentration and skewness in the light distribution, where higher asymmetry values and lower concentration levels signify galaxies that are more irregular and more likely to be undergoing interactions. For the purposes of this study, we focus solely on concentration and asymmetry, while neglecting the smoothness parameter, as discussed in Section \ref{sec:CAS}.

\begin{figure}
    \centering
    \includegraphics[width=0.45\textwidth]{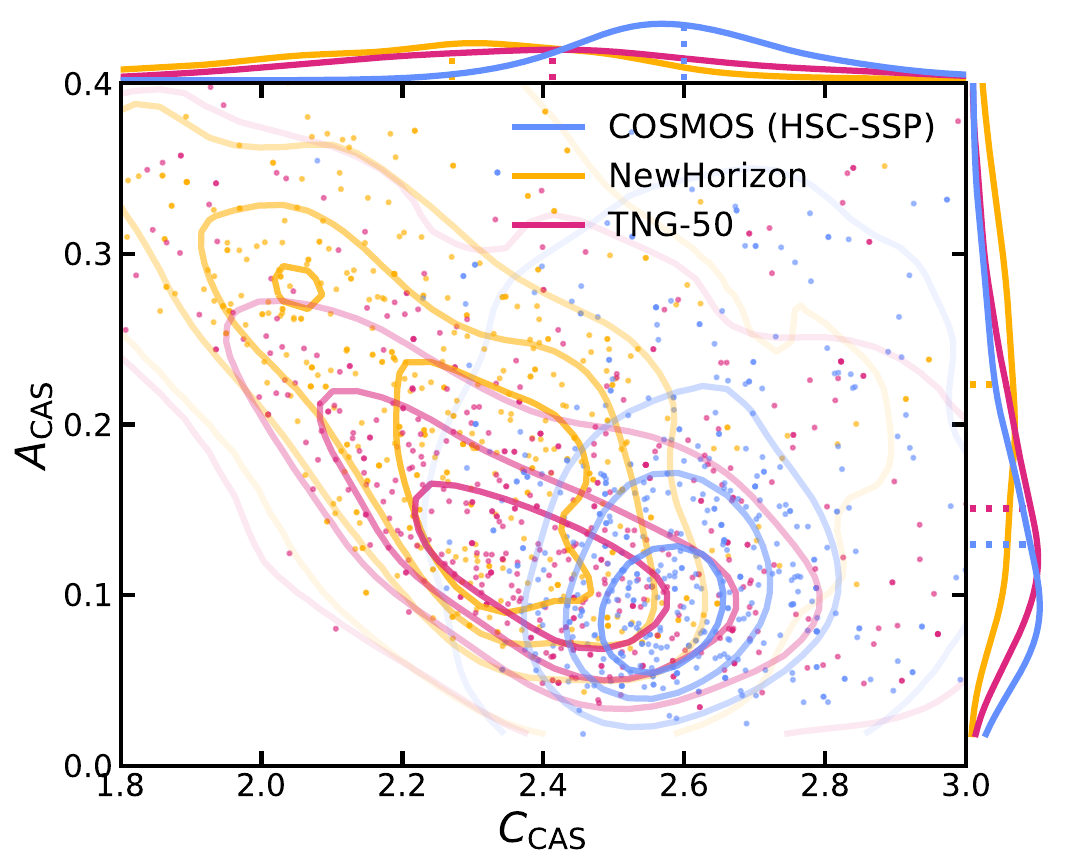}
    \caption{Contour plot showing the 2-d distribution of concentration and asymmetry for COSMOS (blue) and the redshift and mass-matched samples from \textsc{NewHorizon} (yellow) and \textsc{TNG50} (red). Coloured points show a randomly selected sub-sample with the same colour scheme. The sides of the plot show the marginal distribution of each variable with dashed lines indicating the median values. See Figure \ref{fig:CAS_rest} for a plot showing the rest-frame quantities at fixed angular scale.}
    \label{fig:CAS}
\end{figure}

Figure \ref{fig:CAS} shows the 2-d distribution of concentration and asymmetry for COSMOS (blue) and the redshift and mass-matched samples from \textsc{NewHorizon} (yellow) and \textsc{TNG50} (red). The marginal distribution of each parameter is shown along the sides of the plot, with dashed lines indicating the median value of each distribution.

\textsc{NewHorizon} dwarfs are both less concentrated and more asymmetric than the COSMOS dwarfs, indicating that they have less regular shapes and are more morphologically disturbed on average. TNG50 dwarfs have levels of asymmetry similar to COSMOS dwarfs, but their concentrations are closer to those of \textsc{NewHorizon} dwarfs. However, when examining the same quantities in the rest-frame (Figure \ref{fig:CAS_rest}), it is clear that \textsc{TNG50} dwarfs are in fact considerably more concentrated than they appear in the observed-frame, likely even more so than the observed COSMOS dwarfs, whose low-redshift sample exhibit effective radii larger than the PSF FWHM. This again appears to be due to the bias resulting from their smaller sizes. For the \textsc{NewHorizon} dwarfs, the wider range of asymmetry values seen in the observed-frame is still borne out in the rest-frame distribution.

On average, \textsc{NewHorizon} galaxies have asymmetry parameters more consistent with being interacting systems \citep[e.g.][]{Conselice2009}. Given the, on average, more rarefied environments inhabited by \textsc{NewHorizon} dwarfs and the expected decline in merger rate with halo mass \citep[e.g.][]{Fakhouri2010,Martin2021}, it is unlikely that these trends are a genuine result of mergers. Instead, higher asymmetries are more likely due to internal processes which result in a clumpier light distribution.

There are a number of potential explanations for this, including differences in spatial resolution, ISM physics, star formation efficiencies and the way in which SN feedback is modelled and coupled to the ISM, which may result in the suppression stellar clumps and localised bursts of star formation. We discuss these in detail in Section \ref{sec:why}.

\begin{figure}
    \centering
    \includegraphics[width=0.45\textwidth]{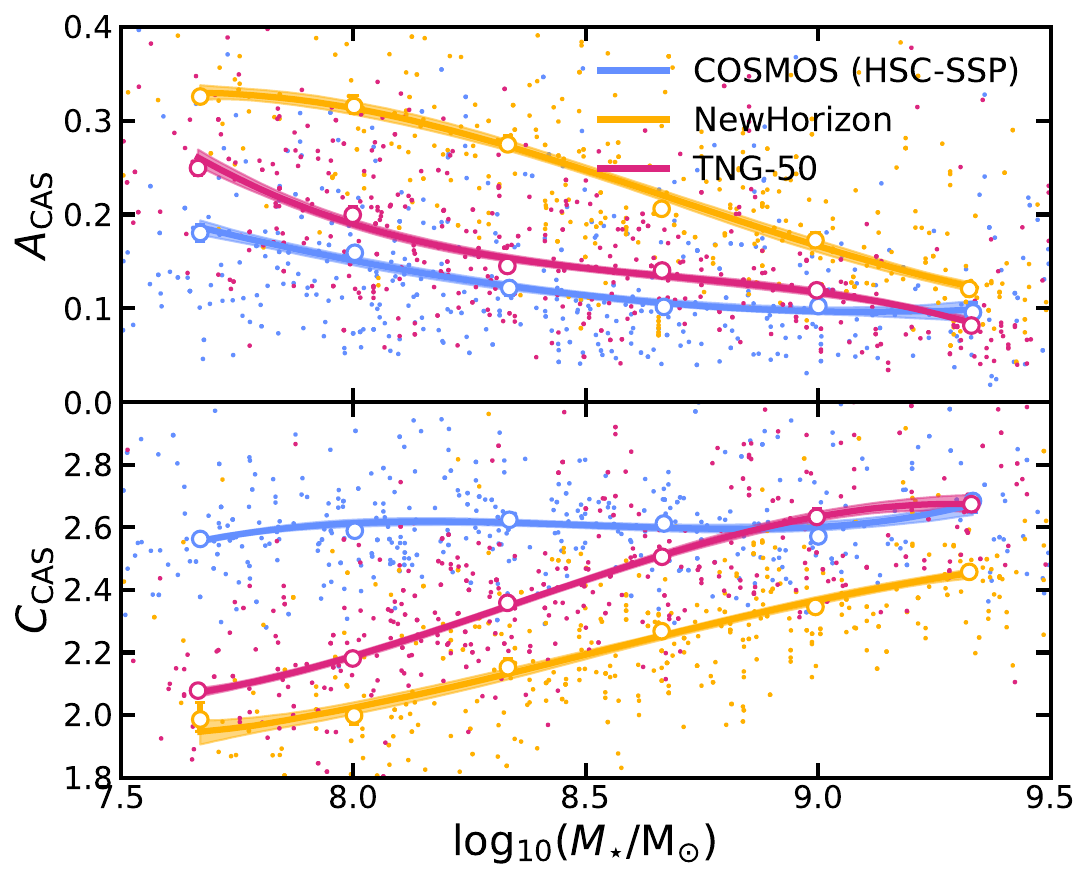}
    \caption{Plots showing the trend in the median asymmetry and concentration as a function of stellar mass for COSMOS (blue) and the redshift and mass-matched samples from \textsc{NewHorizon} (yellow) and \textsc{TNG50} (red). Open circles with error bars show the median and error on the median for individual redshift bins, with filled regions indicating the $1\sigma$ uncertainty. Coloured points show a randomly selected sub-sample with the same colour scheme. See Figure \ref{fig:rest_cas_mass} for a plot showing the rest-frame quantities at fixed angular scale.}
    \label{fig:CAS_mass}
\end{figure}

Figure \ref{fig:CAS_mass} shows the trend in the median asymmetry and concentration values as a function on mass. COSMOS dwarfs exhibit limited changes across the mass range, while \textsc{NewHorizon} and \textsc{TNG50} dwarfs appear more concentrated and more symmetrical towards higher masses. The same trends are observed in the rest-frame in the case of concentration, but there is no clear trend with stellar mass observed for asymmetry in either COSMOS or the simulations.

\begin{figure*}
    \centering
    \includegraphics[width=0.95\textwidth]{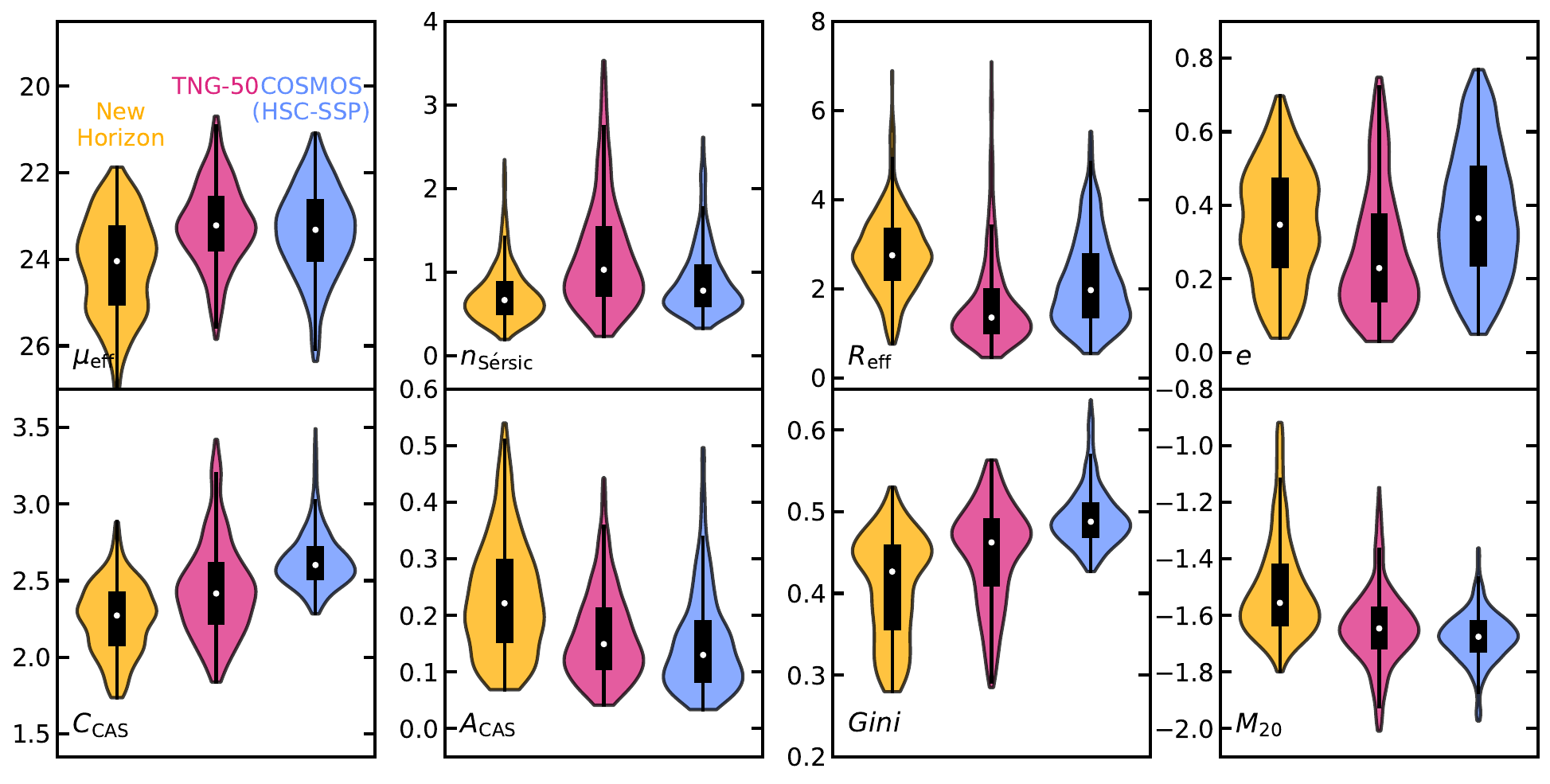}
    \caption{Violin plots summarising the distribution of values discussed throughout Section \ref{sec:results} for \textsc{NewHorizon} (yellow), \textsc{TNG50} (red), and COSMOS (blue). Black box plots overlaid over each violin indicate the inter-quartile range with whiskers representing the extrema, and a white dot indicating the median value of the distribution. See Figure \ref{fig:param_comparison_rest} for a plot showing the rest-frame quantities at fixed angular scale.}
    \label{fig:param_comparison}
\end{figure*}

Figure \ref{fig:param_comparison} provides a comprehensive summary of the parameter distributions discussed in Section \ref{sec:results}. The coloured violin plots illustrate the distributions of each parameter, with yellow, red, and blue violins representing \textsc{NewHorizon}, \textsc{TNG50}, and COSMOS, respectively. To aid visualization, black box plots are overlaid on each violin, indicating the inter-quartile range, whiskers representing the extrema, and a white dot indicating the median value of the distribution. Additionally, a similar plot is presented in Figure \ref{fig:param_comparison_rest}, showing rest-frame quantities at a fixed angular scale, alongside COSMOS dwarfs limited to $z<0.1$.

Notably, both the median values and the overall distributions of \textsc{NewHorizon}, \textsc{TNG50}, and COSMOS dwarfs show significant differences across almost every parameter. These differences suggest substantial variations in the structure of galaxies produced by each simulation in comparison to each other and the observed galaxies. Although there are some differences in the values recovered for some parameters when viewed in the rest-frame (Figure \ref{fig:param_comparison_rest}), large differences between the distributions of simulated and observed dwarfs persist across most parameters. As noted previously in this section, TNG50 dwarfs show significant shifts, particularly for the Gini, $M_{20}$ and CAS parameters, which are not PSF corrected, but these shifts typically do not result in improved agreement with the observations.

Our results in the low-mass regime show some areas of correspondence with the work of \citet{Eisert2024}, who compare intermediate and high-mass TNG50 and TNG100 galaxies to counterparts in HSC observations. We also identify simulated galaxies with apparently unrealistic sizes, S\'ersic profiles and structural parameters, although the ways in which our sample of low-mass TNG50 galaxies diverges from the observed sample differs compared with the higher mass sample of \citet{Eisert2024}. Notably, we find that TNG50 dwarfs are too small compared to their observed counterparts, whereas \citet{Eisert2024} find that more massive galaxies tend to be too large, although anomalous examples in both our low-mass sample and \citet{Eisert2024}'s high mass sample tend to have S\'ersic indices that are too high. Overall, \citet{Eisert2024} find better agreement between observed and simulated TNG50 galaxies. Similarly, other works such as \citet{Ortega2023} show a strong correspondence between the CAS and Gini-$M_{20}$ structural parameters of high-mass TNG50 and KiDS \citep{deJong2013} observed galaxies as a function of mass. We also observe a similarly strong correspondence at the high end of our mass range, where our study has some overlap with  \citet{Ortega2023}. Overall, there appears to be significantly better correspondence in the structural properties of TNG50 and \textsc{NewHorizon} galaxies with their observed counterparts in the higher-mass regime. This correspondence begins to break down as upon entering the lower mass regime ($M_{\star}<10^{9.5}~{\rm M_{\odot}}$), which represents a region where the morphological behaviour and formation mechanisms of galaxies begin to transition \citep[e.g.][]{Lazar2024a}.

\section{Discussion}
\label{sec:discussion}

We have shown that, despite covering a similar range to the observed galaxies in each of the morphological parameters measured in this paper, both \textsc{NewHorizon} and \textsc{TNG50} produce dwarf galaxies whose overall distribution of structural properties are at significant odds with those measured in the COSMOS dwarfs. While \textsc{TNG50} produces more centrally concentrated dwarfs with very small sizes, \textsc{NewHorizon} produces dwarfs at the opposite extreme, having very large sizes and small S\'ersic indices. 

In this section, we discuss possible limitations of our approach and possible physical drivers of the observed differences between the observed and simulated dwarfs.

\subsection{Limitations}

We begin by discussing some of the limits of our comparison between the simulations and between each simulation and the observed COSMOS galaxies.

\subsubsection{Galaxy integrated fluxes and surface brightnesses} As seen in Fig. \ref{fig:sersic}, \textsc{NewHorizon} galaxies have significantly fainter effective surface brightnesses on average than their observed counterparts. \textsc{TNG50} galaxies exhibit similar effective surface brightnesses to COSMOS, but given the fact that they are significantly more compact when compared with their observed counterparts, this implies they still have fainter integrated brightnesses. While we do not rule out a genuine difference in the chemical or star-forming properties of galaxies driving this, there are several choices made when producing our synthetic images that also impact the brightness of the object.

\begin{enumerate}
    \item The choice of the IMF used to produce synthetic images influences both the mass-to-light ratio and the stellar mass loss prescription and can therefore significantly influence the integrated magnitudes of galaxies. \citet{Watkins2025} show that adopting a \citet{Salpeter1955} IMF produces a decrease in the integrated magnitudes of the bluest galaxies of approximately 1~mag with a negligible effect on red galaxies. Given that both \citet{Chabrier2003} and \citet{Salpeter1955} are calibrated to the Milky Way, the adopted IMF may not be entirely appropriate or may vary significantly between galaxies and environments in lower mass regimes \citep[e.g.][]{Weidner2005,Geha2013}. In the case of this study we have tended to make assumptions that result in fainter magnitudes. 
    \item Similarly, our specific choice of SED template, in this case those of \citet{Bruzual2003}, represent another source of uncertainty when comparing simulated galaxies to their observed counterparts because different SED models assume different stellar spectral libraries or stellar evolution tracks.
\end{enumerate}

Given the differences observed in their structural properties, it is likely that there are genuine and significant physical differences between the chemical and star-forming properties of the simulated and observed dwarf galaxies. In the second part of this series of papers, we will revisit the subject of galaxy star formation histories and chemical enrichment.

\subsubsection{Biases due to the PSF}

Due to the compact physical sizes of the galaxies under investigation in this paper, which depending on their redshift, can be close to the FWHM of the HSC PSF, it is important to consider the effect of seeing on our ability to spatially resolve them. In particular, this may lead to biases or significant uncertainties in recovered structural properties. As mentioned previously, we investigated the magnitude of this bias by examining the properties that we recover in the rest-frame at a fixed angular scale.

Despite the potential for bias in these measurements due to the smaller sizes of \textsc{TNG50} galaxies, the fixed angular scale rest-frame quantities clearly demonstrate real and highly significant differences in the structural properties of galaxies between the two simulations. These plots are made available in Appendix \ref{A:restframe} and, while showing similar qualitative trends, \textsc{TNG50} galaxies in particular exhibit notably higher concentrations, S\'ersic indices and notably smaller effective radii, indicating that many of them are compact enough that they become poorly resolved compared with the PSF at higher redshifts. For reference the equivalent physical scale of the average FWHM of the HSC $i$-band PSF at $z=0.1$ and $z=0.2$ is approximately 1.3 and 2.3~kpc respectively, while the median effective radius of a \textsc{TNG50} dwarf is only slightly larger than 1~kpc.

With upcoming large-area surveys such as the Legacy Survey of Space and Time \citep[LSST][]{Ivezic2019,Watkins2024} and higher angular resolution space-based instruments like Euclid \citep{Mellier2024}, significantly larger resolved and mass complete samples of dwarf galaxies will soon be available. This will enable considerably more robust comparison with simulated low-mass galaxies.

\subsection{Physical drivers}
\label{sec:why}

Finally, we discuss some possible physical drivers of the observed differences. A detailed investigation will be undertaken in the second part of this series of papers (Martin et al., in prep). A more comprehensive discussion of environment and baryonic feedback specific to the \textsc{NewHorizon} simulation can be found in \citet{Watkins2025}.

\subsubsection{Environment}

Environmental differences likely contribute to the observed discrepancies. For instance, \citet{Mercado2025} report a similar mismatch in the mass-size relation for low-mass galaxies in the FIREBox simulation, attributing much of the size evolution to environmental effects, particularly their impact on dark matter halo masses. Previous studies using \textsc{NewHorizon} \citep{Jackson2021} and its lower-resolution parent simulation Horizon-AGN \citep{Martin2019} also indicate that the diffuse nature of low-mass galaxies arises from a combination of stellar feedback and environmental influences. Observationally, studies such as \citet{Privon2017} and \citet{Lazar2024b} suggest that the relationship between environment, star formation, and morphology differs significantly between low- and high-mass galaxies.

Environment is expected to influence galaxy properties through both large-scale tidal torques, which shape angular momentum buildup, and smaller-scale interactions with other galaxies, which become more frequent in denser environments. While these processes likely contribute to dwarf galaxy structure, COSMOS, \textsc{TNG50}, and \textsc{NewHorizon} primarily probe `average' (field and group) environments, making it unlikely that environment alone fully explains the observed discrepancies.

As we will show in the second paper of this series (Martin et al., in prep), restricting the \textsc{TNG50} sample--which spans a larger box size and therefore covers somewhat more diverse environments--to the same local density range as \textsc{NewHorizon} reveals some evolution in the morphological and star-forming properties of the galaxies. However, these changes are not sufficient to bring \textsc{TNG50} into clear agreement with either \textsc{NewHorizon} or COSMOS, suggesting that other factors play a more dominant role \citep[e.g.][]{Romeo2020}.

One such factor could be gas resolution, which has been shown to influence galaxy sizes at higher masses. Higher resolution allows for more accurate modelling of angular momentum loss on small scales, leading to more compact galaxies \citep{Chabanier2020}. However, the opposite effect is observed in our case, possibly due to the lower mass regime or the greater importance of other physical processes governing galaxy evolution.

\textsc{NewHorizon} dwarfs appear significantly more morphologically disturbed than their counterparts in COSMOS, exhibiting notably lower concentrations and higher asymmetries. \textsc{TNG50} dwarfs display considerably higher concentrations to those in COSMOS and \textsc{NewHorizon} once the effect of PSF smearing has been taken into account. This could be explained by differences in environment leading to different frequencies of interactions, but could also be the result of physical differences in the galaxies themselves, such as differing halo or stellar mass profiles, which might influence how firmly material is retained within the potential well of the galaxy halo \citep[e.g.][]{Martin2019,Jackson2021,Martin2024}. Given the considerably larger sizes and less concentrated stellar mass profiles of the \textsc{NewHorizon} dwarfs, it is plausible that this could lead to more pronounced irregularities and stronger tidal features, as material can be more easily moved or liberated from the galaxy during interactions. As discussed later, another likely driver of higher asymmetries may lie in the differences in implemented sub-grid physics, which could lead to burstier star formation and more clumpy stellar distributions.

\subsubsection{ISM, star formation and feedback}

Beyond differences in numerical resolution, \textsc{NewHorizon} and \textsc{TNG50} adopt significantly different treatments of the ISM and star formation physics, leading to contrasting galaxy properties. \textsc{TNG50} assumes an ISM model that relies on an over-pressurised medium driven by internal feedback effects with the \cite{Springel2003} model. Their star formation follows a Schmidt law with a constant SF efficiency ($\simeq 6.5\%$) for gas densities larger than $n>0.13\,\rm cm^{-3}$, and is combined to a hydrodynamical decoupling of the SN feedback from the star-forming gas densities so that the energy is released in diffuse gas.

In contrast, \textsc{NewHorizon} does not assume an effective equation of state for the ISM, allowing star formation to occur in denser gas ($n > 10\,\rm cm^{-3}$) with a variable efficiency that ranges from nearly zero to several tens of percent \citep{Dubois2021}. Here, SN feedback is coupled to the surrounding gas in a local fashion, resulting in more effective feedback on cold star-forming gas.

Those different approaches to the ISM gas physics and stellar feedback have important consequences on the structure of the ISM: \textsc{NewHorizon} produces a multiphase ISM in which dense, star-forming clumps are efficiently disrupted by SNe, leading to a more dynamic, turbulent medium. In contrast, \textsc{TNG50} generates a smoother ISM, where star-forming gas is more long-lived and less susceptible to episodic feedback-driven disruption.

The more bursty and variable star formation in \textsc{NewHorizon} likely drives stronger episodic outflows that preferentially eject central, low-angular-momentum gas \citep[e.g.][]{Namin2024}. The loss of this gas is likely to suppress star formation in the central regions and result in a more diffuse and extended stellar distribution. Such feedback-regulated gas redistribution also helps explain the more irregular morphologies and clumpy substructures seen in \textsc{NewHorizon} galaxies.

In contrast, the ISM model in \textsc{TNG50} promotes smoother, more continuous feedback that lacks the violent, episodic ejections required to effectively disrupt central regions. Consequently, low-angular-momentum gas accumulates in the galaxy centres, fuelling central star formation and producing more compact, concentrated structures \citep[e.g.][]{Almeida2024}. While strong stellar feedback can quench star formation, it has also been shown to lead to overly compact galaxies. Additionally, the inclusion of MHD in \textsc{TNG50} may further enhance central compactness by regulating gas turbulence and outflows \citep[e.g.][]{2023MartinAlvarez}.

Although the models used in \textsc{NewHorizon} and \textsc{TNG50} produce galaxy structural properties that differ significantly from observations, some integrated properties, such as the galaxy stellar mass function \citep[e.g.][]{Dubois2021}, have been shown to be consistent. \citet{Wright2024} have further demonstrated that simulations with different feedback implementations can yield similar integrated properties through distinct mechanisms. This highlights that reproducing integrated properties alone does not ensure the underlying feedback processes are physically accurate. More detailed, resolved 2-d properties—such as those presented in this paper and by \citet{Watkins2025}—therefore provide a more stringent test of the simulations' accuracy and greater power to constrain underlying physical mechanisms.

\section{Summary}

\label{sec:summary}

In this paper we have investigated the structural properties of dwarf galaxies by analysing data from both the \textsc{NewHorizon} and \textsc{TNG50} simulations, in addition to ultra-deep HSC-SSP observations of the COSMOS field. Our approach involves first producing synthetic images of simulated galaxies matching the mass and redshift distribution of observed dwarfs in the COSMOS field and then injecting these synthetic images into realistic backgrounds taken from the HSC-SSP data. This methodology enables direct and robust comparison of the structural properties of observed and simulated dwarf galaxies. We use the S\'ersic profile fitting and non-parametric, Gini, $M_{20}$, asymmetry and concentration parameters to compare the distribution of observed and simulated galaxies. We summarise our findings below.

\begin{enumerate}
\item \textit{Simulated and observed dwarf galaxies show distinct structural differences.}
\textsc{NewHorizon} and TNG50 produce galaxies with structural properties that diverge significantly, lying to opposite extremes of the observed COSMOS dwarfs. \textsc{NewHorizon} produces diffuse, extended galaxies with shallow S\'ersic indices, while TNG50 yields highly compact, concentrated galaxies with steep Sérsic indices. Observed COSMOS dwarfs span an intermediate range of structural properties, suggesting that neither simulation fully captures the structural diversity of real dwarf galaxies.
\\
\item \textit{Non-parametric measures also indicate discrepancies.} Gini, $M_{20}$ and CAS measurements show that \textsc{NewHorizon} galaxies have more uneven, and clumpy light distributions which are more likely to be linked to more bursty star formation processes than environmental processes. TNG50 galaxies exhibit smoother but excessively concentrated light profiles (in the rest-frame) compared with observed COSMOS dwarfs.
\\
\item \textit{Structural trends with stellar mass reveal differences between simulations and observations.} In \textsc{NewHorizon} and TNG50, galaxy sizes, concentrations, and asymmetries evolve with mass but follow significantly different trends. \textsc{NewHorizon} dwarfs remain too diffuse even towards the high end of the mass range ($M_{\star}=10^{9.5}~{\rm M_{\odot}}$), although their S\'ersic indices remain consistently shallow in agreement with the COSMOS dwarfs. TNG50 dwarfs exhibit a strong trend toward higher concentrations and steeper S\'ersic indices towards higher masses, which aligns poorly with observations. COSMOS dwarfs, in contrast, show a much less pronounced trend with stellar mass, with relatively stable structural and morphological properties across the mass range.
\\
\item \textit{At the high end of the mass range, \textsc{NewHorizon} and TNG50 galaxies begin to better reproduce the general structural trends seen in COSMOS dwarfs.} Most S\'ersic and structural properties of \textsc{NewHorizon} and TNG50 dwarfs fall into better agreement with COSMOS at higher masses. TNG50 galaxies, however, remain somewhat too compact and retain S\'ersic indices significantly steeper than observed COSMOS or \textsc{NewHorizon} dwarfs. These findings are consistent with previous studies \citep[e.g.][]{Eisert2024}, which have shown that TNG50 tends to achieve better agreement with observations in the higher mass regime, contrasting with the larger discrepancies observed at lower masses in this study.
\\
\item \textit{Observational effects alone cannot explain simulation discrepancies.}
While observational factors like seeing introduce biases, repeating our analysis in the rest-frame confirms intrinsic differences in simulated galaxy properties. In particular, many TNG50 galaxies are smeared out by the PSF due to their extremely compact intrinsic sizes, meaning they are even more compact than they appear in the observed-frame, while \textsc{NewHorizon} galaxies remain diffuse and asymmetric even in the absence of observational biases.
\\
\item \textit{Differences in simulation physics offer valuable insights into the drivers of galaxy evolution.} The distinct structural properties of \textsc{NewHorizon} and TNG50 reflect the underlying differences in their physical models, including their ISM physics, star formation prescriptions and feedback implementations. The pronounced variations in the structural properties of simulated dwarfs in this mass regime underscore the heightened sensitivity of dwarf galaxy evolution to these processes. New observatories, such as the Vera C. Rubin Observatory and Euclid, will facilitate much more robust comparisons with simulations of low-mass galaxies, enabling stronger constraints on the physical mechanisms that shape their evolution.

\end{enumerate}

We have shown that low-mass galaxy populations produced by two state-of-the-art simulations have structural properties that are significantly at odds with both each other and observed galaxies in the same mass regime. This disparity may offer insight into the underlying physics shaping galaxy evolution, particularly the physics of star formation and stellar feedback, to which galaxy properties are especially sensitive in this regime. In the second part of this series of papers (Martin et al., in prep), we will investigate the correlation between galaxy morphology and galaxy star formation history in this regime.


\section*{Acknowledgements}

G.~M is thankful to Taysun Kimm, Nina Hatch and Emmanuele Contini for fruitful discussions.

G.~M acknowledges support from the UK STFC under grant ST/X000982/1.

AEW acknowledges support from the STFC under grant ST/X001318/1.

S.~K.~Y. acknowledges support from the National Research Foundation of Korea (2020R1A2C3003769; RS-2022-NR070872).

D.K. acknowledges support from the National Research Foundation of Korea (NRF) grant funded by the Korean government(MSIT) (No. NRF-2022R1C1C2004506).

SK, IL and AEW acknowledge support from the STFC (grant numbers ST/Y001257/1 and ST/X001318/1). SK also acknowledges a Senior Research Fellowship from Worcester College Oxford.

Part of this work was carried out by G.~M during a Balzan Visiting Fellowship held at the Department of Physics, University of Oxford.

The \textsc{NewHorizon} simulation was undertaken with HPC resources of CINES under the allocations  c2016047637, A0020407637 and A0070402192 by Genci, KSC-2017-G2-0003 by KISTI, and as a “Grand Challenge” project granted by GENCI on the AMD Rome extension of the Joliot Curie supercomputer at TGCC. A large data transfer was supported by KREONET which is managed and operated by KISTI.

The \textsc{IllustrisTNG} simulations were undertaken with compute time awarded by the Gauss Centre for Supercomputing (GCS) under GCS Large-Scale Projects GCS-ILLU and GCS-DWAR on the GCS share of the supercomputer Hazel Hen at the High Performance Computing Center Stuttgart (HLRS), as well as on the machines of the Max Planck Computing and Data Facility (MPCDF) in Garching, Germany.

This work has made use of the Infinity cluster on which the \textsc{\textsc{NewHorizon} }simulation was post-processed, hosted by the Institut d'Astrophysique de Paris. We warmly thank S.~Rouberol for running it smoothly.

This research is part of the Spin(e) ANR-13-BS05-0005 http://cosmicorigin.org), Segal ANR-19-CE31-0017 (http://secular-evolution.org).

This research made use of \textit{Photutils}, an Astropy package for detection and photometry of astronomical sources \citep{Bradley2022}.

The Hyper Suprime-Cam (HSC) collaboration includes the astronomical communities of Japan and Taiwan, and Princeton University. The HSC instrumentation and software were developed by the National Astronomical Observatory of Japan (NAOJ), the Kavli Institute for the Physics and Mathematics of the Universe (Kavli IPMU), the University of Tokyo, the High Energy Accelerator Research Organization (KEK), the Academia Sinica Institute for Astronomy and Astrophysics in Taiwan (ASIAA), and Princeton University. Funding was contributed by the FIRST program from the Japanese Cabinet Office, the Ministry of Education, Culture, Sports, Science and Technology (MEXT), the Japan Society for the Promotion of Science (JSPS), Japan Science and Technology Agency (JST), the Toray Science Foundation, NAOJ, Kavli IPMU, KEK, ASIAA, and Princeton University. 

This paper makes use of software developed for Vera C. Rubin Observatory. We thank the Rubin Observatory for making their code available as free software at http://pipelines.lsst.io/.

This paper is based in part on data collected at the Subaru Telescope and retrieved from the HSC data archive system, which is operated by the Subaru Telescope and Astronomy Data Center (ADC) at NAOJ. Data analysis was in part carried out with the cooperation of Center for Computational Astrophysics (CfCA), NAOJ. We are honored and grateful for the opportunity of observing the Universe from Maunakea, which has cultural, historical and natural significance in Hawaii.

\section*{Data Availability}

The simulation data analysed in this paper were provided by the Horizon collaboration and the Illustris project. \textsc{NewHorizon} data will be shared on request to the corresponding author, with the permission of the Horizon collaboration or may be requested from \url{https://new.horizon-simulation.org/data.html}. \textsc{Illustris TNG50} data is publicly available and can be accessed at \url{https://tng-project.org/}.

The third public release of HSC-SSP data is publicly available and can be accessed at \url{https://hsc-release.mtk.nao.ac.jp/}.




\bibliographystyle{mnras}
\bibliography{paper_mnras} 



\appendix

\section{Depth of HSC-SSP Cosmos field}

\label{A:depth}

In this section we present estimates of the $i$-band depth of the HSC-SSP \texttt{deepCoadd} data in the Cosmos field.

A $256\times256$ pixel bin third-order sky correction task is performed on every exposure before co-addition. This can lead to systematics in background subtraction \citep[][and references therein]{Watkins2024}.

objects smaller than this scale (which is the case for all of the objects we consider here) should not suffer from over-subtraction.

We split the central pointing of the Cosmos field into a $9\times9$ element grid, measuring the limiting surface brightness within each square by first masking all pixels with detections and then further splitting each square into a grid with elements of $256\times256$ pixels (the same scale on which the sky correction is performed) and calculating the median standard deviation across each grid element. 

We then convert to a limiting surface brightness according to the following equation from \citep{Roman2020}:
\begin{equation}
    \mu_{lim}(n\sigma,\Omega\times\Omega) = -2.5\,{\rm log_{10}}\left(\frac{n\sigma}{pix\times\Omega}\right)+Z_{p}\,,
\end{equation}
where $\sigma$ is the measured standard deviation of the image counts, $\Omega$ is the box length in arcseconds, $pix$ is the pixel size in arcseconds and $Z_{p}$ is the zero-point (which is set to 27 mag for all \texttt{deepCoadd} images).(10x10, 3 sigma)

\begin{figure}
    \centering
    \includegraphics[width=0.45\textwidth]{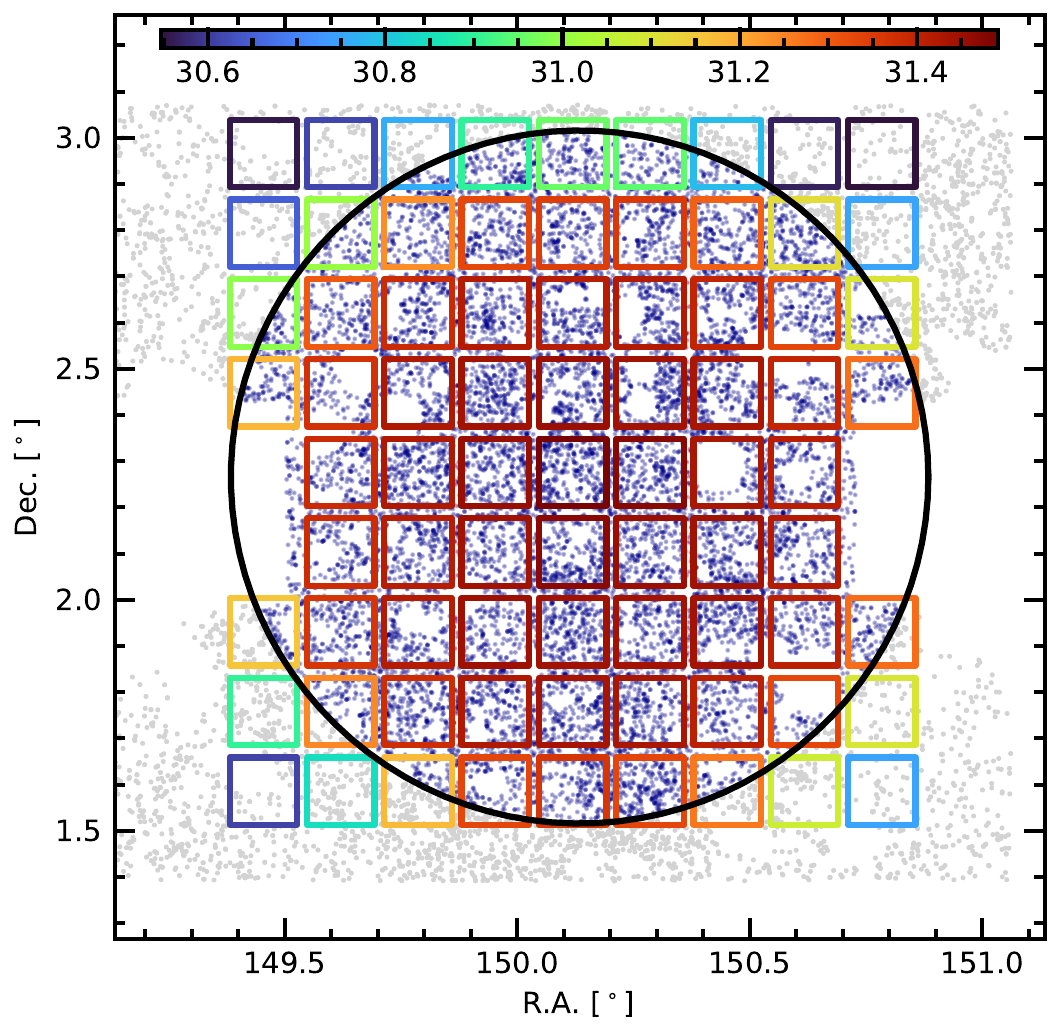}
    \caption{Scatter plot showing the coordinates of COSMOS2020 galaxies within the central $1.5^{\circ}$ of the COSMOS field (bounded by a black circle). The sample used in this study is indicated in blue, while objects lying outside of the central $1.5^{\circ}$ are shown in grey. Overplotted is a $9\times9$ grid, where the colour of the lines indicates the median depth within the region bounded by each grid element. Colours correspond to the colour bar at the top of the plot and indicate the depth as a limiting surface brightness $\mu^{\rm lim}_{i}(3\sigma$, $10^{\prime\prime}\times10^{\prime\prime})$ measured in mag~arcsec$^{-2}$.}
    \label{fig:SB_Cosmos}
\end{figure}

Figure \ref{fig:SB_Cosmos} illustrates the measured COSMOS depth within the central $1.5^\circ$ field of view. The $i$-band surface brightness limit, $\mu^{\rm lim}_{i}(3\sigma, 10^{\prime\prime}\times10^{\prime\prime})$, exceeds 31~mag~arcsec$^{-2}$, with mean and median values of 31.35~mag~arcsec$^{-2}$ and 31.41~mag~arcsec$^{-2}$, respectively.

To evaluate object recovery under realistic conditions, we inject synthetic galaxies--sourced from the \textsc{NewHorizon} and \textsc{TNG50} datasets--into HSC-SSP \texttt{deepCoadd} images. These galaxies are inserted into empty regions of the sky to minimize systematic effects from source confusion. Specifically, injection positions are chosen within the central COSMOS field such that they are at least 20 pixels away from any detected source. This separation is sufficiently large to avoid confusion with nearby objects, yet remains well below the $256\times256$ pixel bin size used for sky correction, ensuring that potential systematics from sky background estimation near other sources are captured. We note that galaxies projected close to extended or heavily clustered sources might still be affected; however, such cases are expected to increase scatter in our measurements rather than introduce systematic biases.

After injecting the synthetic galaxies, we rerun the detection and segmentation algorithm to produce an updated segmentation map that includes the injected objects. A detection is deemed successful if the centroid of the closest detected object lies within 10 pixels of the injected galaxy’s center (i.e., half the minimum injection separation).

To estimate the equivalent limiting surface brightness, we also perform a separate set of detections under idealized conditions—using only Gaussian random noise—across a grid of limiting surface brightnesses between 29 and 32~mag~arcsec$^{-2}$. We define ${\rm MAX}(\mu_{\rm lim, idealised})$ as the faintest limiting surface brightness at which a given object remains detectable after noise injection, and $f_{\rm HSC, detected}$ as the fraction of objects injected into the HSC \texttt{deepCoadd} imaging that are detected for a given bin of ${\rm MAX}(\mu_{\rm lim, idealised})$.

\begin{figure}
    \centering
    \includegraphics[width=0.45\textwidth]{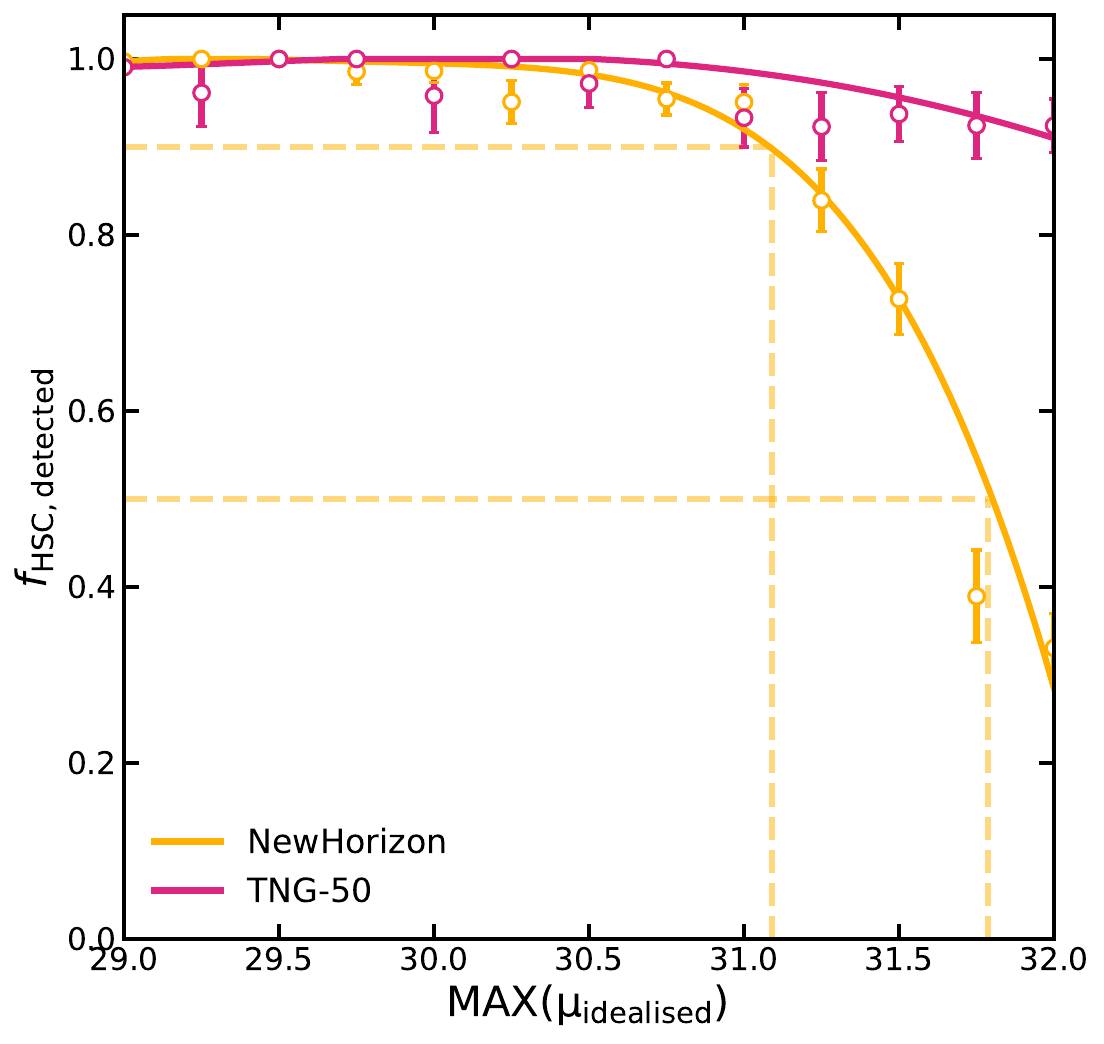}
    \caption{Detection efficiency of synthetic galaxy injections in HSC-SSP \texttt{deepCoadd} images as a function of the idealized limiting surface brightness, ${\rm MAX}(\mu_{\rm lim, idealised})$.}
    \label{fig:SB_Cosmos_injection}
\end{figure}

We find that the \textsc{NewHorizon} sample is 90 per cent and 50 per cent complete up to limiting surface brightnesses, $\mu^{\rm lim}_{i}(3\sigma$, $10^{\prime\prime}\times10^{\prime\prime}$, of 31.1 and 31.8~mag~arcsec$^{-2}$ respectively. In the case of \textsc{TNG50} galaxies, which are significantly more compact, the sample is over 90 per cent complete beyond 32~mag~arcsec$^{-2}$. These values appear consistent with our previous limiting surface brightness estimate of $\sim31.4$~mag~arcsec${-2}$. Note that we would not expect exact correspondence, as the limiting surface brightness measured directly from the background noise and the limiting surface brightness at which an object is detected are not equivalent measurements.

\section{Rest-frame quantities}
\label{A:restframe}

In this Section, we present `rest-frame' versions of the plots from Section \ref{sec:results}. S\'ersic, CAS and gini-$M_{20}$ parameters are measured in the rest-frame with a fixed angular scale of $1~{\rm kpc/{\prime\prime}}$.

\begin{figure}
    \centering
    \includegraphics[width=0.45\textwidth]{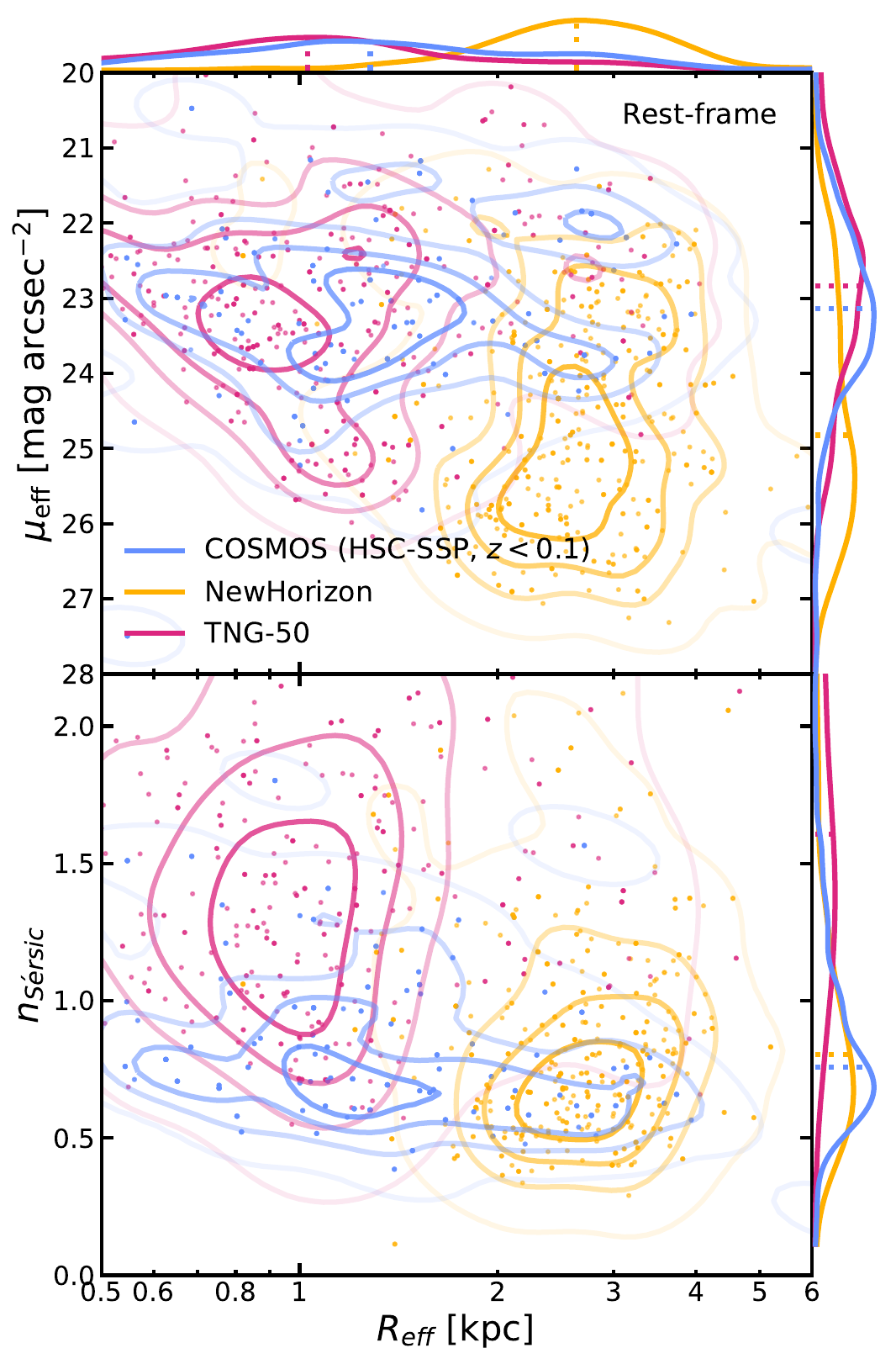}
    \caption{Contour plots showing the 2-d distribution of the rest-frame measured surface brightness at the effective radius (top panel) and S\'ersic index (bottom panel) as a function of effective radius  for the redshift and mass-matched samples from \textsc{NewHorizon} (yellow) and \textsc{TNG50} (red) with a $z<0.1$ restricted sub-sample of COSMOS (blue). Coloured points show a randomly selected sub-sample with the same colour scheme. The sides of each panel show the marginal distribution of each variable with dashed lines indicating the median values.}
    \label{fig:sersic_rest}
\end{figure}

\begin{figure}
    \centering
    \includegraphics[width=0.45\textwidth]{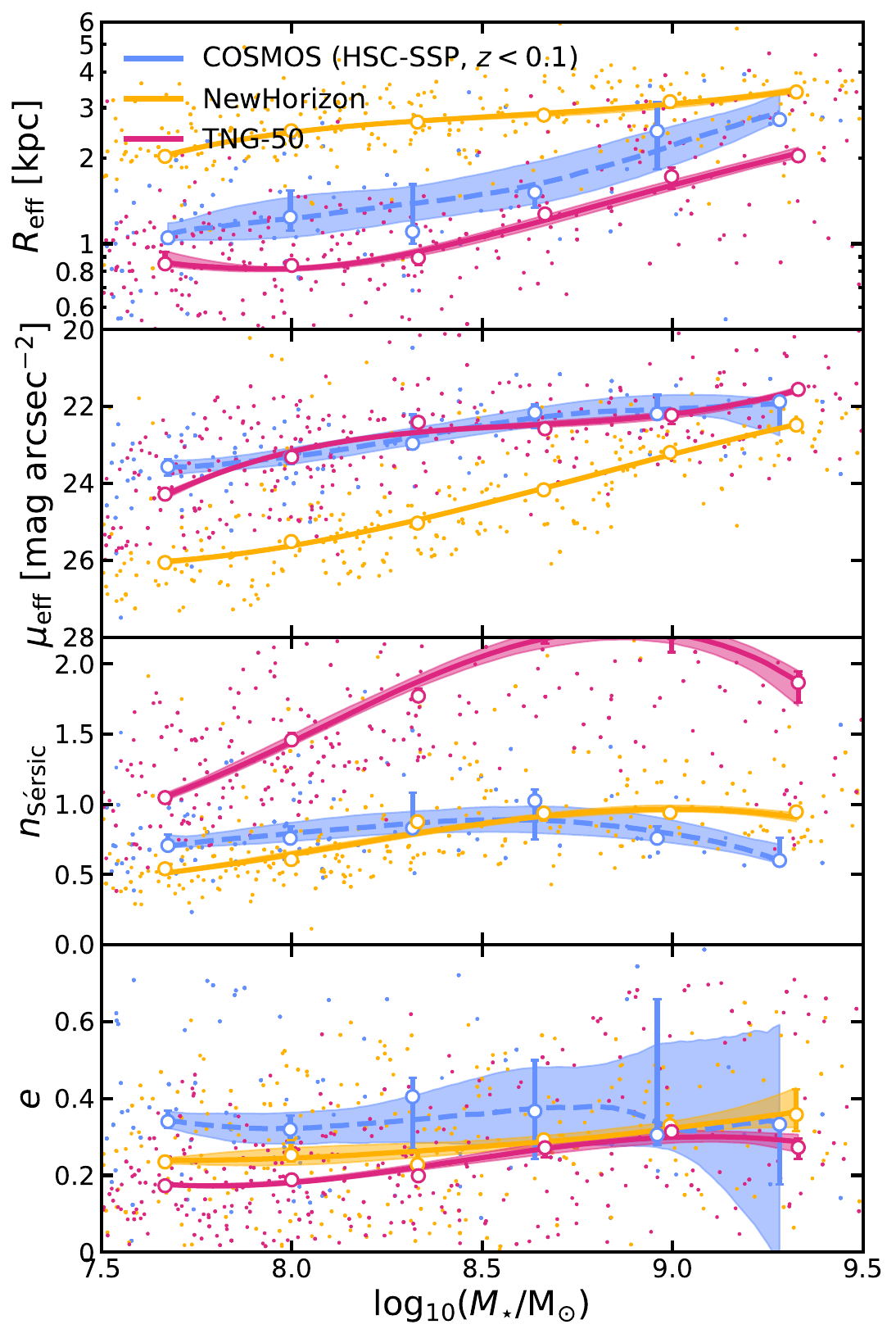}
    \caption{The evolution of the median rest-frame effective radius, surface brightness at the effective radius, S\'ersic index and projected ellipticity as a function of stellar mass for a $z<0.1$ restricted sub-sample of COSMOS (blue) and the redshift and mass-matched samples from \textsc{NewHorizon} (yellow) and \textsc{TNG50} (red). Open circles with error bars show the median and error on the median for individual redshift bins, with filled regions indicating the $1\sigma$ uncertainty. Coloured points show a randomly selected sub-sample with the same colour scheme.}
    \label{fig:rest_sersic_mass}
\end{figure}

\begin{figure}
    \centering
    \includegraphics[width=0.45\textwidth]{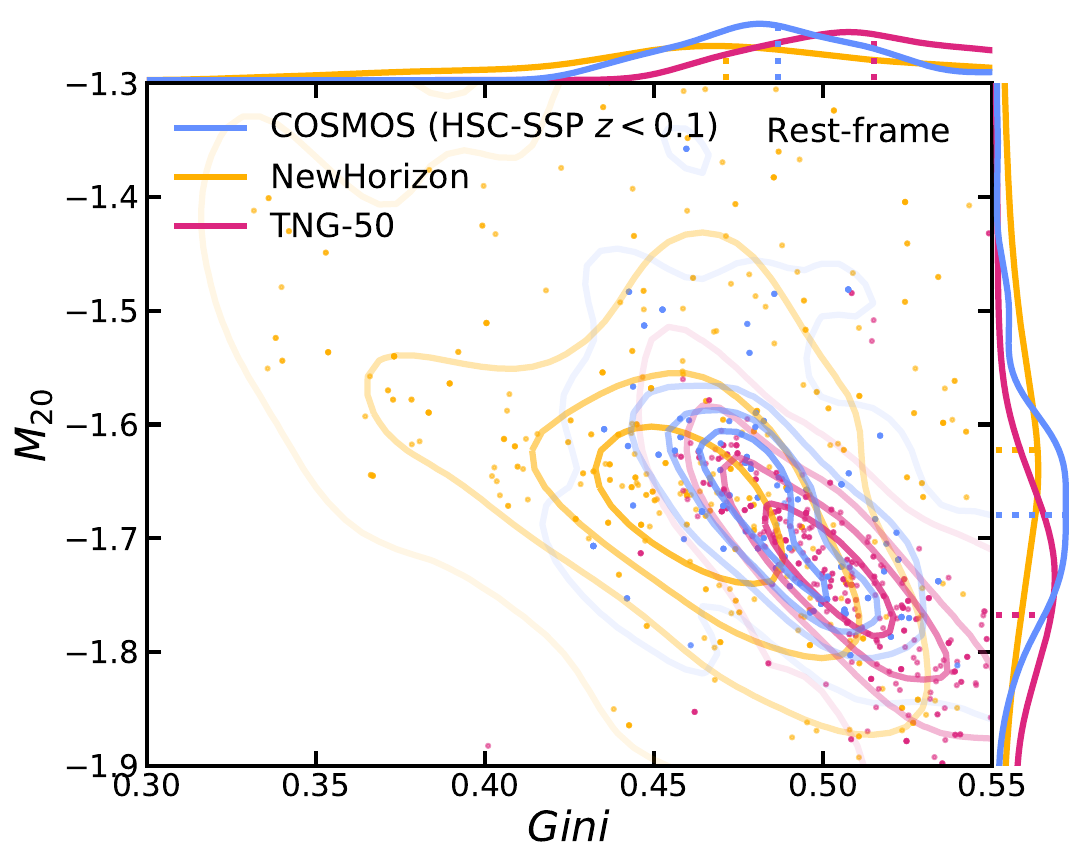}
    \caption{Contour plot showing the 2-d distribution of the rest-frame measured Gini and $M_{20}$ for the redshift and mass-matched samples from \textsc{NewHorizon} (yellow) and \textsc{TNG50} (red) with a $z<0.1$ restricted sub-sample of COSMOS (blue). Coloured points show a randomly selected sub-sample with the same colour scheme. The sides of the plot show the marginal distribution of each variable with dashed lines indicating the median values.}
    \label{fig:giniM20_rest}
\end{figure}

\begin{figure}
    \centering
    \includegraphics[width=0.45\textwidth]{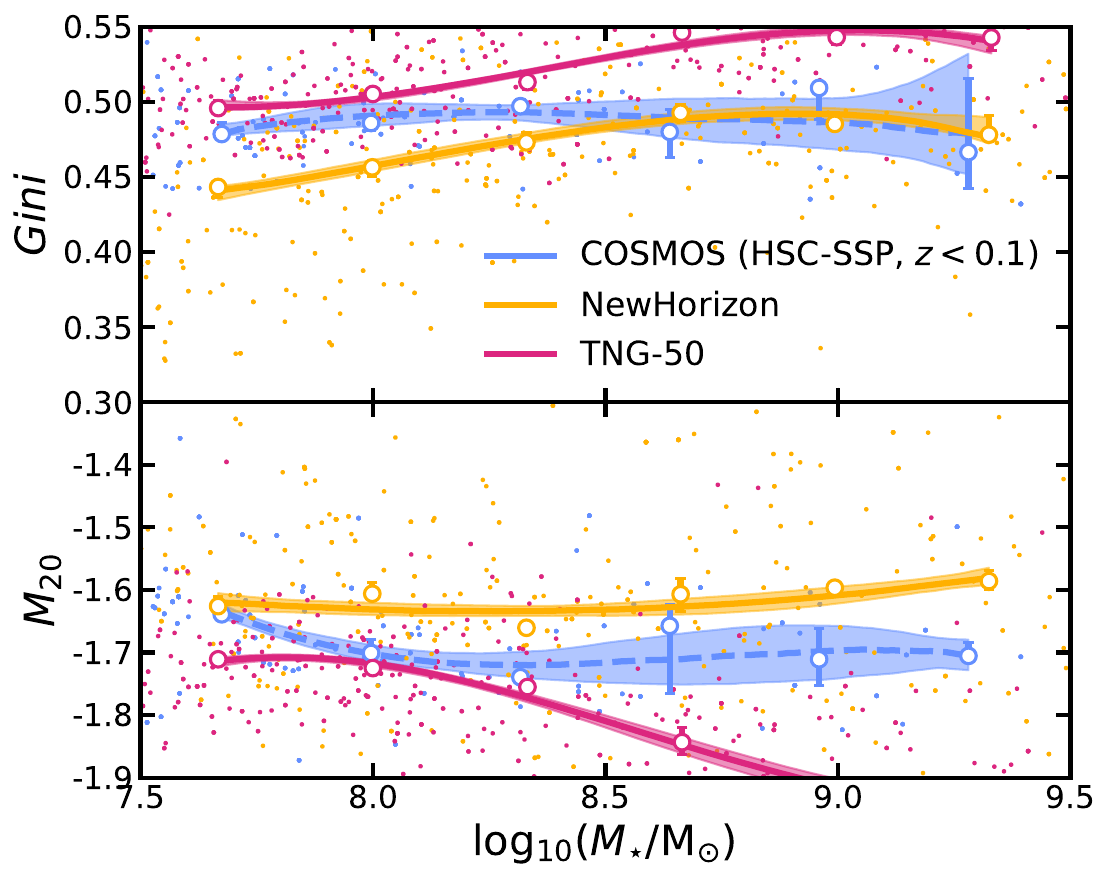}
    \caption{Plots showing the evolution of the median rest-frame measured gini and $M_{20}$ as a function of stellar mass for a $z<0.1$ restricted sub-sample of COSMOS and the redshift and mass-matched samples from \textsc{NewHorizon} (yellow) and \textsc{TNG50} (red). Open circles with error bars show the median and error on the median for individual redshift bins, with filled regions indicating the $1\sigma$ uncertainty. Coloured points show a randomly selected sub-sample with the same colour scheme.}
    \label{fig:rest_ginim20_mass}
\end{figure}

\begin{figure}
    \centering
    \includegraphics[width=0.45\textwidth]{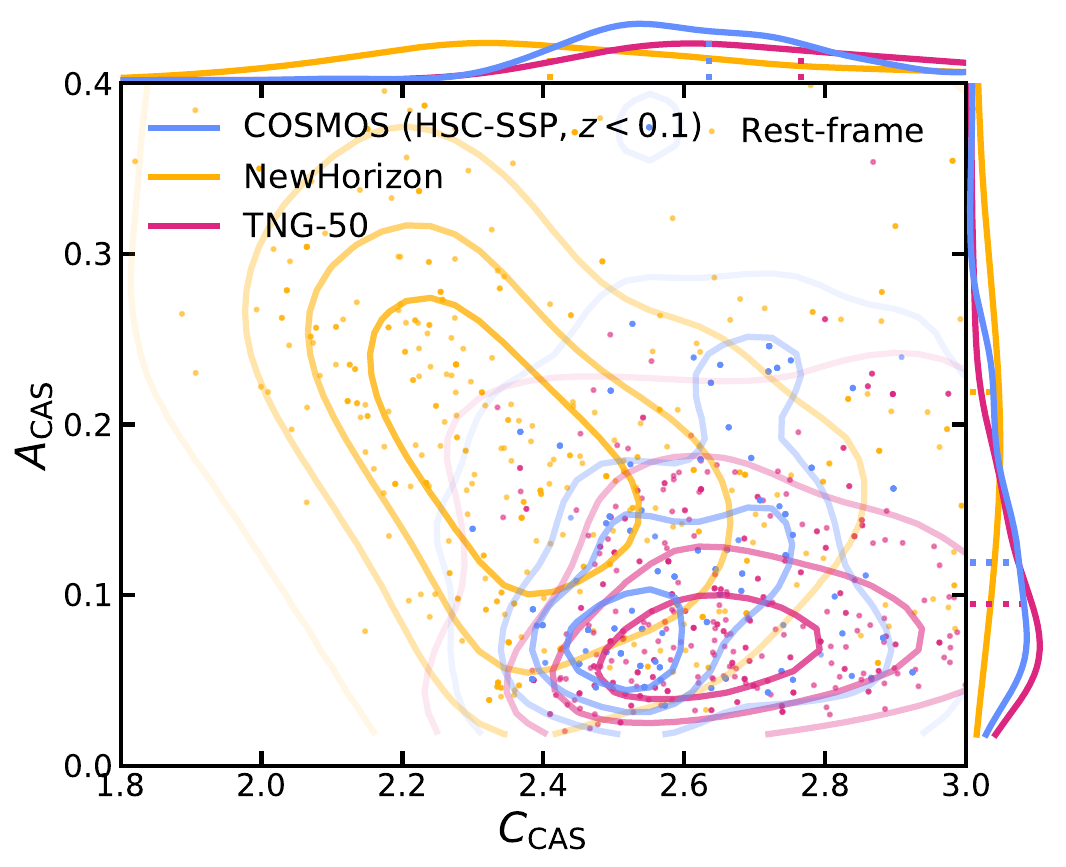}
    \caption{Contour plot showing the 2-d distribution of the rest-frame measured concentration and asymmetry for the redshift and mass-matched samples from \textsc{NewHorizon} (yellow) and \textsc{TNG50} (red) with a $z<0.1$ restricted sub-sample of COSMOS (blue). Coloured points show a randomly selected sub-sample with the same colour scheme. The sides of the plot show the marginal distribution of each variable with dashed lines indicating the median values.}
    \label{fig:CAS_rest}
\end{figure}

\begin{figure}
    \centering
    \includegraphics[width=0.45\textwidth]{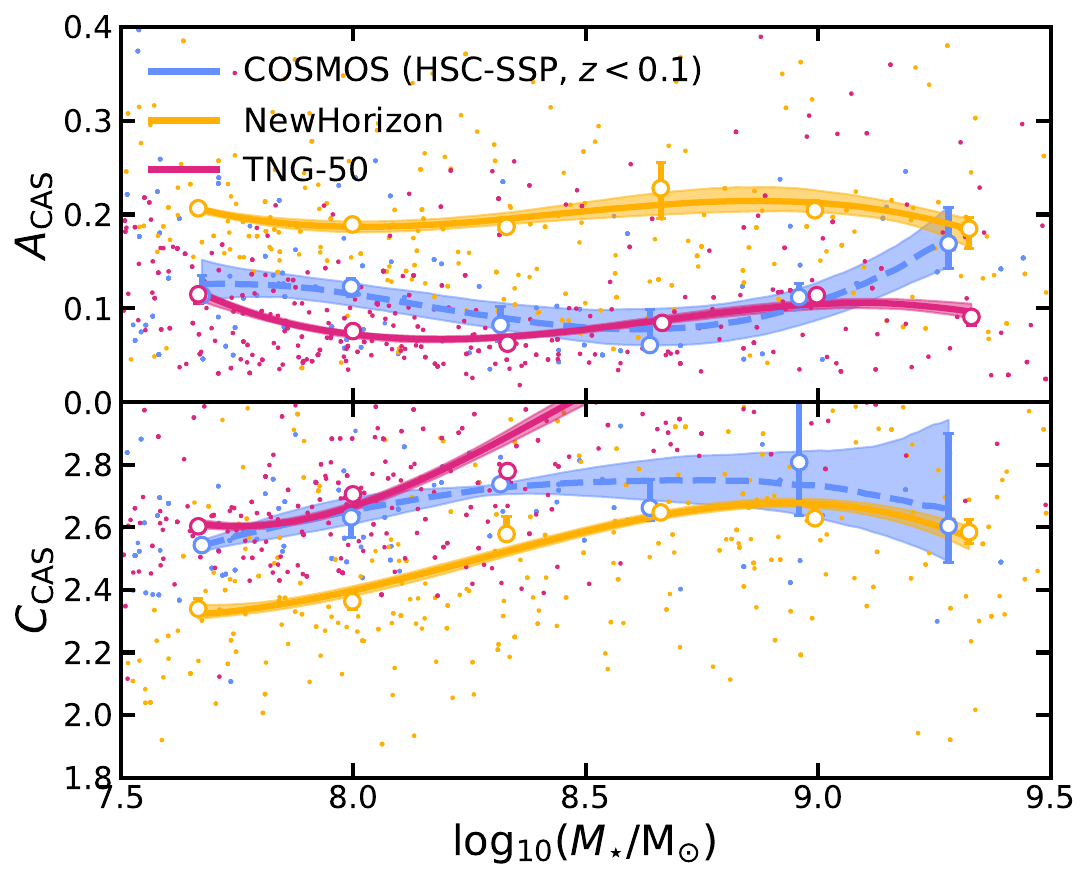}
    \caption{Plots showing the evolution of the median rest-frame measured asymmetry and concentration as a function of stellar mass for a $z<0.1$ restricted sub-sample of COSMOS and the redshift and mass-matched samples from \textsc{NewHorizon} (yellow) and \textsc{TNG50} (red). Open circles with error bars show the median and error on the median for individual redshift bins, with filled regions indicating the $1\sigma$ uncertainty. Coloured points show a randomly selected sub-sample with the same colour scheme.}
    \label{fig:rest_cas_mass}
\end{figure}

\begin{figure*}
    \centering
    \includegraphics[width=0.95\textwidth]{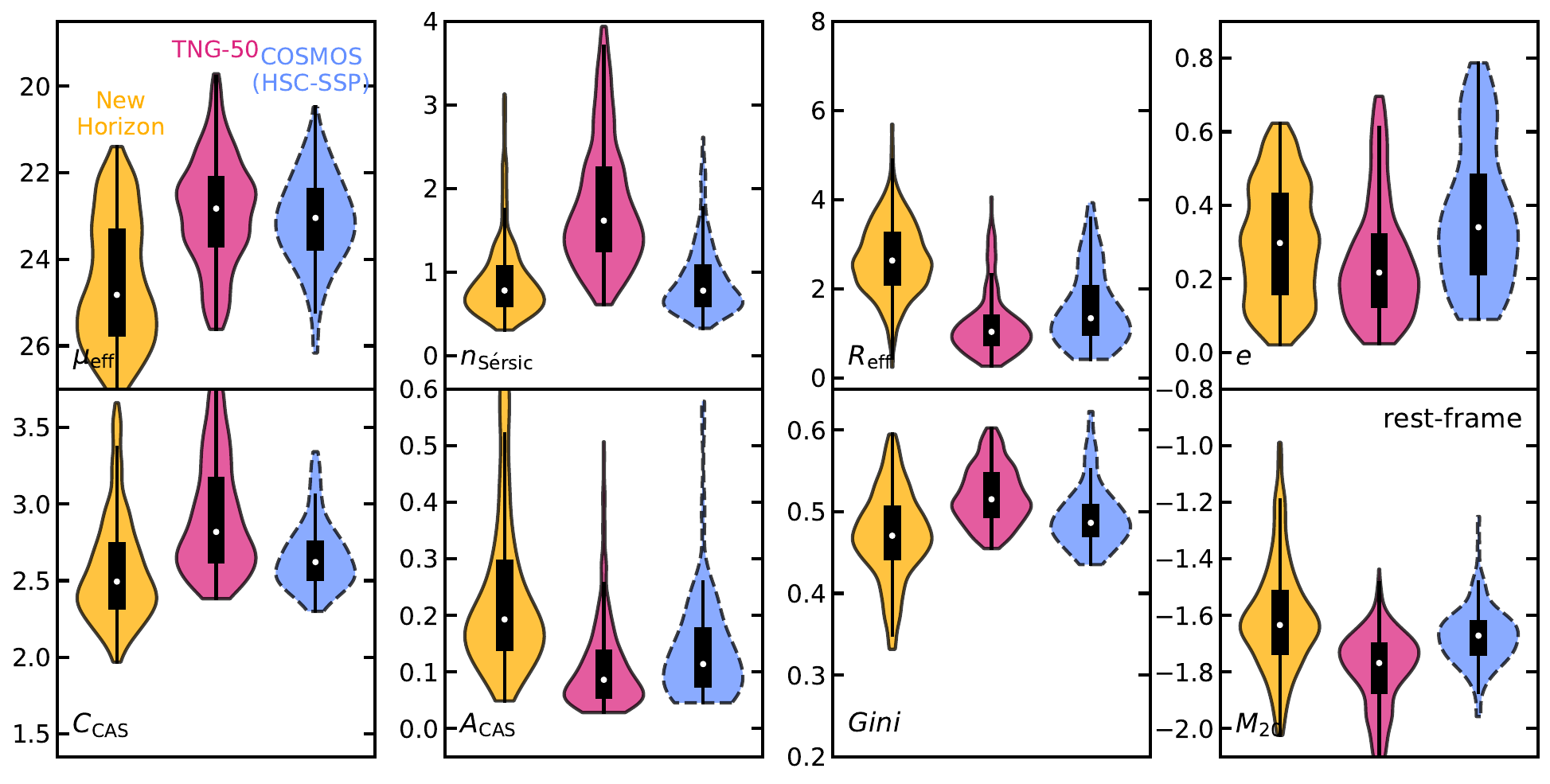}
    \caption{Violin plots summarising the distribution of rest-frame values discussed throughout Section \ref{sec:results} for \textsc{NewHorizon} (yellow), \textsc{TNG50} (red), and a $z<0.1$ restricted sub-sample of COSMOS (blue). Black box plots overlaid over each violin indicate the inter-quartile range with whiskers representing the extrema, and a white dot indicating the median value of the distribution.}
    \label{fig:param_comparison_rest}
\end{figure*}




\bsp	
\label{lastpage}
\end{document}